\documentclass[a4paper,UKenglish,cleveref, autoref, thm-restate]{lipics-v2021}

\title{Proving Quantum Programs Correct (Extended Version)}

\author{Kesha Hietala}{University of Maryland, College Park, USA \and \url{https://www.cs.umd.edu/~kesha/}}{kesha@cs.umd.edu}{https://orcid.org/0000-0002-2724-0974}{}

\author{Robert Rand}{University of Chicago, USA \and \url{https://people.cs.uchicago.edu/~rand/}}{rand@uchicago.edu}{https://orcid.org/0000-0001-6842-5505}{}

\author{Shih-Han Hung}{University of Maryland, College Park, USA}{shung@cs.umd.edu}{https://orcid.org/0000-0003-3410-7466}{}

\author{Liyi Li}{University of Maryland, College Park, USA}{liyili2@umd.edu}{https://orcid.org/0000-0001-8184-0244}{}

\author{Michael Hicks}{University of Maryland, College Park, USA \and \url{http://www.cs.umd.edu/~mwh/}}{mwh@cs.umd.edu}{https://orcid.org/0000-0002-2759-9223}{}

\authorrunning{K. Hietala, R. Rand, S. Hung, L. Li, and M. Hicks} 

\Copyright{Kesha Hietala, Robert Rand, Shih-Han Hung, Liyi Li, and Michael Hicks} 



\category{} 

\relatedversion{}

\supplement{}
\supplementdetails[]{Software}{https://github.com/inQWIRE/SQIR}

\funding{This material is based upon work supported by the U.S. Department of Energy, Office of Science, Office of Advanced Scientific Computing Research, Quantum Testbed Pathfinder Program under Award Number DE-SC0019040 and the Air Force Office of Scientific Research under Grant No. FA95502110051. }

\acknowledgements{We thank Yuxiang Peng for ongoing contributions to the \sqir codebase and pointing out a bug in our original specification for QPE. We thank Xiaodi Wu for discussions about \sqir and follow-on projects.}

\nolinenumbers 

 \hideLIPIcs  

\EventEditors{Liron Cohen and Cezary Kaliszyk}
\EventNoEds{2}
\EventLongTitle{12th International Conference on Interactive Theorem Proving (ITP 2021)}
\EventShortTitle{ITP 2021}
\EventAcronym{ITP}
\EventYear{2021}
\EventDate{June 29--July 1, 2021}
\EventLocation{Rome, Italy (Virtual Conference)}
\EventLogo{}
\SeriesVolume{193}
\ArticleNo{7}

\usepackage{stmaryrd}
\usepackage{xcolor}
\usepackage{colortbl}

\usepackage[utf8]{inputenc}

\usepackage{amsmath}
\usepackage{hyperref}

\usepackage{subcaption}
\usepackage{wrapfig}

\usepackage{xspace}
\usepackage{float}
\usepackage{balance}

\usepackage{textcomp} 

\usepackage{cleveref} 

\usepackage{wrapfig}
\usepackage{caption}

\usepackage{longtable}
\usepackage{tabu}

\usepackage{mathtools} 
\usepackage{xfrac} 

\usepackage[qm,braket]{qcircuit}
\newcommand{\ketbra}[2]{\ket{#1}\!\bra{#2}}

\newcommand{\sqir}{\textsc{sqir}\xspace}
\newcommand{\qwire}{\ensuremath{\mathcal{Q}\textsc{wire}}\xspace}
\newcommand{\qbricks}{\ensuremath{\mathcal{Q}\textsc{bricks}}\xspace}

\newcommand{\voqc}{\textsc{voqc}\xspace}

\usepackage{tikz}

\newcommand{\N}{\ensuremath{\mathbb{N}}\xspace}

\newcommand{\CC}{\ensuremath{\mathbb{C}}\xspace}

\usepackage{tikz} 
\usetikzlibrary{%
  arrows,%
  shapes.misc,
  shapes.geometric, 
  shapes.arrows,%
  shapes.callouts,
  shapes.gates.logic.US,
  chains,%
  matrix,%
  positioning,
  scopes,%
  decorations.pathmorphing,
  decorations.text,
  decorations.pathreplacing, 
  shadows%
}
\tikzset{ machine/.style={
    rectangle,
    minimum width=25mm,
    minimum height=18mm,
    text width=24mm,
    align=center,
    very thick,
    draw=black,
    color=black,
    fill=white,
  }
}

\DeclarePairedDelimiter\abs{\lvert}{\rvert}
\DeclarePairedDelimiter\norm{\lVert}{\rVert}

\makeatletter
\let\oldabs\abs
\def\abs{\@ifstar{\oldabs}{\oldabs*}}
\let\oldnorm\norm
\def\norm{\@ifstar{\oldnorm}{\oldnorm*}}
\makeatother

\makeatletter
\DeclareRobustCommand{\vardivision}{%
  \mathbin{\mathpalette\@vardivision\relax}%
}
\newcommand{\@vardivision}[2]{%
  \reflectbox{$\m@th\smallsetminus$}%
}
\makeatother

\usepackage{booktabs}

\usepackage{alltt}
\usepackage{listings,lstcoq}
\definecolor{ltblue}{rgb}{0,0.4,0.4}
\definecolor{dkblue}{rgb}{0,0.1,0.6}
\definecolor{dkgreen}{rgb}{0,0.35,0}
\definecolor{dkviolet}{rgb}{0.3,0,0.5}
\definecolor{dkred}{rgb}{0.5,0,0}
\newcommand{\code}[1]{{\small\texttt{#1}}}

\newcommand{\denote}[1]{\llbracket #1 \rrbracket\xspace}
\newcommand{\pdenote}[1]{\{\hspace{-0.2em}| #1 |\hspace{-0.2em}\}}

\usepackage{newunicodechar}
\let\Alpha=A
\let\Beta=B
\let\Epsilon=E
\let\Zeta=Z
\let\Eta=H
\let\Iota=I
\let\Kappa=K
\let\Mu=M
\let\Nu=N
\let\Omicron=O
\let\omicron=o
\let\Rho=P
\let\Tau=T
\let\Chi=X

\newunicodechar{Α}{\ensuremath{\Alpha}}
\newunicodechar{α}{\ensuremath{\alpha}}
\newunicodechar{Β}{\ensuremath{\Beta}}
\newunicodechar{β}{\ensuremath{\beta}}
\newunicodechar{Γ}{\ensuremath{\Gamma}}
\newunicodechar{γ}{\ensuremath{\gamma}}
\newunicodechar{Δ}{\ensuremath{\Delta}}
\newunicodechar{δ}{\ensuremath{\delta}}
\newunicodechar{Ε}{\ensuremath{\Epsilon}}
\newunicodechar{ε}{\ensuremath{\epsilon}}
\newunicodechar{ϵ}{\ensuremath{\varepsilon}}
\newunicodechar{Ζ}{\ensuremath{\Zeta}}
\newunicodechar{ζ}{\ensuremath{\zeta}}
\newunicodechar{Η}{\ensuremath{\Eta}}
\newunicodechar{η}{\ensuremath{\eta}}
\newunicodechar{Θ}{\ensuremath{\Theta}}
\newunicodechar{θ}{\ensuremath{\theta}}
\newunicodechar{ϑ}{\ensuremath{\vartheta}}
\newunicodechar{Ι}{\ensuremath{\Iota}}
\newunicodechar{ι}{\ensuremath{\iota}}
\newunicodechar{Κ}{\ensuremath{\Kappa}}
\newunicodechar{κ}{\ensuremath{\kappa}}
\newunicodechar{Λ}{\ensuremath{\Lambda}}
\newunicodechar{λ}{\ensuremath{\lambda}}
\newunicodechar{Μ}{\ensuremath{\Mu}}
\newunicodechar{μ}{\ensuremath{\mu}}
\newunicodechar{Ν}{\ensuremath{\Nu}}
\newunicodechar{ν}{\ensuremath{\nu}}
\newunicodechar{Ξ}{\ensuremath{\Xi}}
\newunicodechar{ξ}{\ensuremath{\xi}}
\newunicodechar{Ο}{\ensuremath{\Omicron}}
\newunicodechar{ο}{\ensuremath{\omicron}}
\newunicodechar{Π}{\ensuremath{\Pi}}
\newunicodechar{π}{\ensuremath{\pi}}
\newunicodechar{ϖ}{\ensuremath{\varpi}}
\newunicodechar{Ρ}{\ensuremath{\Rho}}
\newunicodechar{ρ}{\ensuremath{\rho}}
\newunicodechar{ϱ}{\ensuremath{\varrho}}
\newunicodechar{Σ}{\ensuremath{\Sigma}}
\newunicodechar{σ}{\ensuremath{\sigma}}
\newunicodechar{ς}{\ensuremath{\varsigma}}
\newunicodechar{Τ}{\ensuremath{\Tau}}
\newunicodechar{τ}{\ensuremath{\tau}}
\newunicodechar{Υ}{\ensuremath{\Upsilon}}
\newunicodechar{υ}{\ensuremath{\upsilon}}
\newunicodechar{Φ}{\ensuremath{\Phi}}
\newunicodechar{φ}{\ensuremath{\phi}}
\newunicodechar{ϕ}{\ensuremath{\varphi}}
\newunicodechar{Χ}{\ensuremath{\Chi}}
\newunicodechar{χ}{\ensuremath{\chi}}
\newunicodechar{Ψ}{\ensuremath{\Psi}}
\newunicodechar{ψ}{\ensuremath{\psi}}
\newunicodechar{Ω}{\ensuremath{\Omega}}
\newunicodechar{ω}{\ensuremath{\omega}}

\newunicodechar{ℕ}{\ensuremath{\mathbb{N}}}
\newunicodechar{∅}{\ensuremath{\emptyset}}

\newunicodechar{∙}{\ensuremath{\bullet}}
\newunicodechar{≈}{\ensuremath{\approx}}
\newunicodechar{≅}{\ensuremath{\cong}}
\newunicodechar{≡}{\ensuremath{\equiv}}
\newunicodechar{≤}{\ensuremath{\le}}
\newunicodechar{≥}{\ensuremath{\ge}}
\newunicodechar{≠}{\ensuremath{\neq}}
\newunicodechar{∀}{\ensuremath{\forall}}
\newunicodechar{∃}{\ensuremath{\exists}}
\newunicodechar{±}{\ensuremath{\pm}}
\newunicodechar{∓}{\ensuremath{\pm}}
\newunicodechar{·}{\ensuremath{\cdot}}
\newunicodechar{⋯}{\ensuremath{\cdots}}
\newunicodechar{…}{\ensuremath{\ldots}}
\newunicodechar{∷}{~\mathrel{:\!\!\!:}~}
\newunicodechar{×}{\ensuremath{\times}}
\newunicodechar{∞}{\ensuremath{\infty}}
\newunicodechar{→}{\ensuremath{\to}}
\newunicodechar{←}{\ensuremath{\leftarrow}}
\newunicodechar{⇒}{\ensuremath{\Rightarrow}}
\newunicodechar{↦}{\ensuremath{\mapsto}}
\newunicodechar{↝}{\ensuremath{\leadsto}}
\newunicodechar{∨}{\ensuremath{\vee}}
\newunicodechar{∧}{\ensuremath{\wedge}}
\newunicodechar{⊢}{\ensuremath{\vdash}}
\newunicodechar{⊣}{\ensuremath{\dashv}}
\newunicodechar{∣}{\ensuremath{\mid}}
\newunicodechar{∈}{\ensuremath{\in}}
\newunicodechar{⊆}{\ensuremath{\subseteq}}
\newunicodechar{⊂}{\ensuremath{\subset}}
\newunicodechar{∪}{\ensuremath{\cup}}
\newunicodechar{⋓}{\ensuremath{\Cup}}
\newunicodechar{∉}{\ensuremath{\not\in}}
\newunicodechar{√}{\ensuremath{\sqrt}}

\newunicodechar{⊸}{\ensuremath{\multimap}}
\newunicodechar{⊗}{\ensuremath{\otimes}}
\newunicodechar{⨂}{\ensuremath{\bigotimes}}
\newunicodechar{⊕}{\ensuremath{\oplus}}
\newunicodechar{〈}{\ensuremath{\langle}}
\newunicodechar{⟨}{\ensuremath{\langle}}
\newunicodechar{⟩}{\ensuremath{\rangle}}
\newunicodechar{〉}{\ensuremath{\rangle}}
\newunicodechar{¡}{\ensuremath{\upsidedownbang}}
\newunicodechar{∘}{\ensuremath{\circ}}
\newunicodechar{†}{\ensuremath{\dagger}}
\newunicodechar{⊤}{\ensuremath{\top}}
\newunicodechar{⊥}{\ensuremath{\bot}}

\newunicodechar{〚}{\ensuremath{\llbracket}}
\newunicodechar{〛}{\ensuremath{\rrbracket}}

\usepackage{etoolbox} 
\newtoggle{comments}
\toggletrue{comments}

\iftoggle{comments}{
  \newcommand{\fixme}[1]{\textbf{\textcolor{red}{[ Fixme: #1]}}}
  \newcommand{\todo}[1]{\textbf{\textcolor{green}{[ TODO: #1 ]}}}
  \newcommand{\rnr}[1]{\textbf{\textcolor{blue}{[ Robert: #1 ]}}}
  \newcommand{\mwh}[1]{\textbf{\textcolor{olive}{[ Mike: #1 ]}}}
  \newcommand{\khh}[1]{\textbf{\textcolor{orange}{[ Kesha: #1 ]}}}
  \newcommand{\shh}[1]{\textbf{\textcolor{purple}{[ Shih-Han: #1 ]}}}
  
  \newcommand{\oth}[2]{\textbf{\textcolor{red}{[ #1: #2 ]}}}
}{
  \newcommand{\fixme}[1]{}
  \newcommand{\todo}[1]{}
  \newcommand{\rnr}[1]{}
  \newcommand{\mwh}[1]{}  
  \newcommand{\khh}[1]{}
  \newcommand{\shh}[1]{}
  \newcommand{\xwu}[1]{}
  \newcommand{\oth}[2]{}
}

\newtoggle{submission}
\togglefalse{submission}

\iftoggle{submission}{
  \newcommand{\aref}[1]{\Cref{#1} of the extended version of this paper}
}{
  \newcommand{\aref}[1]{\Cref{#1}}
}

\usepackage{capt-of}

\begin{document}

\maketitle

\begin{abstract}

As quantum computing progresses steadily from theory into practice,
programmers will face a common problem:  How can they be sure that their code does what they intend it to do?  
This paper presents encouraging results in the application of mechanized proof to the domain of quantum programming in the context of the \sqir development.  
It verifies the correctness of a range of a quantum algorithms including Grover's algorithm and quantum phase estimation, a key component of Shor's algorithm.
In doing so, it aims to highlight both the successes and challenges of formal verification in the quantum context and motivate the theorem proving community to target quantum computing as an application domain.

\end{abstract}

\section{Introduction}

Quantum computers are fundamentally different from the ``classical'' computers we have been programming since the development of the ENIAC in 1945. This difference includes a layer of complexity introduced by quantum mechanics: Instead of a deterministic function from inputs to outputs, a quantum program is a function from inputs to a \emph{superposition} of outputs, a notion that generalizes probabilities. As a result, quantum programs are strictly more expressive than probabilistic programs and even harder to get right. While we can test the output of a probabilistic program by comparing its observed distribution to the desired one, doing the same on a quantum computer can be prohibitively expensive and may not fully describe the underlying quantum state. 

This challenge for quantum programming is an opportunity for formal methods. We can use formal methods to \emph{prove}, in advance, that the code implementing a quantum algorithm does what it should for all possible inputs and configurations.

In prior work~\cite{VOQC}, we developed a formally verified optimizer for quantum programs (\voqc), implemented and proved correct in the Coq proof assistant~\cite{coq}. \voqc transforms programs written in \sqir, a \emph{small quantum intermediate representation}. While we designed \sqir to be a compiler intermediate representation, we quickly realized that it was not so different from languages used to write \emph{source} quantum programs, and that the design choices that eased proving optimizations correct could ease proving source programs correct, too. 

To date, we have proved the correctness of implementations of a number of quantum algorithms, including quantum teleportation, Greenberger–Horne–Zeilinger (GHZ) state preparation~\cite{Greenberger1989}, the Deutsch-Jozsa algorithm~\cite{deutsch1992rapid}, Simon's algorithm~\cite{Simon1994}, the quantum Fourier transform (QFT), quantum phase estimation (QPE), and Grover's algorithm~\cite{Grover1996}. QPE is a key component of Shor's prime-factoring algorithm~\cite{Shor94}, today's best-known, most impactful quantum algorithm, with Grover's algorithm for unstructured search being the second. Our implementations can be extracted to code that can 
be executed on quantum hardware or simulated classically,
depending on the problem size and hardware limitations. 

While \sqir was first introduced as part of \voqc, this paper offers two new contributions.
First, it presents a detailed discussion of how \sqir's design supports proofs of correctness. After presenting background on quantum computing (\Cref{sec:background}) and reviewing \sqir (\Cref{sec:sqir}), \Cref{sec:design} discusses key elements of \sqir's design and compares and contrasts them to design decisions made in the related tools \qwire~\cite{Paykin2017}, \qbricks~\cite{qbricks}, and the Isabelle implementation of quantum Hoare logic~\cite{Liu2019}.
\sqir's overall benefit over these tools is its flexibility, supporting multiple semantics and approaches to proof. 
As a second contribution, this paper presents the code, formal specification, and proof sketch of Grover's algorithm, QFT, and QPE, which are the most sophisticated algorithms that we have verified so far (\Cref{sec:examples}).
We comment on the proofs of simpler algorithms in \aref{app:general-verification}.
We believe there is ripe opportunity for further application of formal methods to quantum computing and we hope this paper, and our work on \sqir, paves the way for new research; we sketch open problems in \Cref{sec:conclusions}.

\sqir is implemented in just over 3500 lines of Coq, with an additional 3700 lines of example \sqir programs and proofs; it is freely available on Github.\footnote{\url{https://github.com/inQWIRE/SQIR}}

\section{Background}
\label{sec:background}

We begin with a light background on quantum computing; for a full treatment we recommend the standard text on the subject~\cite{NCbook}. 

\subsection{Quantum States}

A quantum state consists of one or more \emph{quantum bits}. A quantum bit (or \emph{qubit}) can be expressed as a two dimensional vector $\begin{psmallmatrix} \alpha \\ \beta \end{psmallmatrix}$ such that $|\alpha|^2 + |\beta|^2 = 1$. 
The $\alpha$ and $\beta$ are called \emph{amplitudes}. We frequently write this vector as $\alpha\ket{0} + \beta\ket{1}$ where $\ket{0} = \begin{psmallmatrix} 1 \\ 0 \end{psmallmatrix}$ and $\ket{1} = \begin{psmallmatrix} 0 \\ 1 \end{psmallmatrix}$ are \emph{basis states}. When both $\alpha$ and $\beta$ are non-zero, we can think of the qubit as being ``both 0 and 1 at once,'' a.k.a. a \emph{superposition}. For example, $\frac{1}{\sqrt{2}}(\ket{0} + \ket{1})$ is an equal superposition of $\ket{0}$ and $\ket{1}$.

We can join multiple qubits together by means of the \emph{tensor product} ($\otimes$) from linear algebra. For convenience, we write $\ket{i} \otimes \ket{j}$ as $\ket{ij}$ for $i,j \in \{0,1\}$; we may also write $\ket{k}$ where $k \in \mathbb{N}$ is the decimal interpretation of bits $ij$.
We use $\ket{\psi}$ to refer to an arbitrary quantum state. Sometimes a multi-qubit state cannot be expressed as the tensor of individual qubits; such states are called \emph{entangled}. One example is the state $\frac{1}{\sqrt{2}}(\ket{00} + \ket{11})$, known as a \emph{Bell pair}.

\subsection{Quantum Programs}
\label{sec:qprograms}

\begin{figure}[t]
    \captionsetup[subfigure]{justification=centering}
\begin{minipage}[b]{.25\textwidth}
  \small
  \Qcircuit @C=0.5em @R=0.5em {
    \lstick{\ket{0}} & \gate{H} & \ctrl{1} & \qw & \qw \\
    \lstick{\ket{0}} & \qw & \targ & \ctrl{1} & \qw \\
    \lstick{\ket{0}} & \qw & \qw & \targ & \qw
    }
\subcaption{Quantum circuit}
\end{minipage}
\begin{minipage}[b]{.2\textwidth}
\begin{coq}
H 0; 
CNOT 0 1; 
CNOT 1 2
\end{coq}
\subcaption{\sqir assembly}
\end{minipage}
\begin{minipage}[b]{.5\textwidth}
\begin{coq}
Fixpoint ghz (n : nat) : ucom base n :=
  match n with
  | 0 => I 0
  | 1 => H 0
  | S n' => ghz n'; CNOT (n'-1) n'
  end.
\end{coq}
\subcaption{\sqir meta-program}
\end{minipage}
\vspace{1em}

\begin{minipage}[b]{.34\textwidth}
\begin{coq}
box (x,y,z) =>
  gate x     <- H x;
  gate (x,y) <- CNOT (x,y);
  gate (y,z) <- CNOT (y,z);
  output (x,y,z).
\end{coq}
\subcaption{\qwire}
\end{minipage}
\begin{minipage}[b]{.33\textwidth}
\begin{coq}
SEQ(SEQ(PAR(H, PAR(I, I)), 
        PAR(CNOT, I)), 
    PAR(I, CNOT))
\end{coq}
\subcaption{\qbricks-DSL}
\end{minipage}
\begin{minipage}[b]{.26\textwidth}
\begin{coq}
q1 := H q1;
q1,q2 := CNOT q1,q2;
q2,q3 := CNOT q2,q3
\end{coq}
\subcaption{QWhile}
\end{minipage}
\vspace{1em}

\begin{minipage}[b]{\textwidth}
\centering
$(I \otimes CNOT) \times (CNOT \otimes I) \times (H \otimes I \otimes I)$
\subcaption{Matrix expression}
\end{minipage}
\caption{Example quantum program: GHZ state preparation.}
\label{fig:circuit-example}
\end{figure}
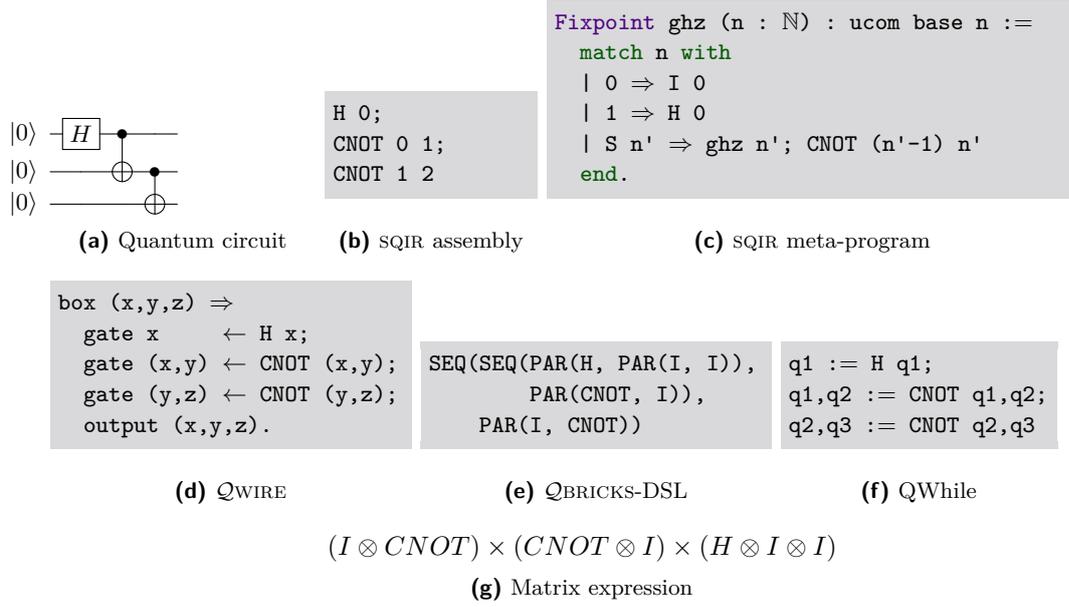

Quantum programs are composed of a series of \emph{quantum operations}, each of which acts on a subset of qubits in the quantum state. In the standard presentation, quantum programs are expressed as \emph{circuits}, as shown in \Cref{fig:circuit-example}(a). In these circuits, each horizontal wire represents a qubit and boxes on these wires indicate quantum operations, or \emph{gates}. The circuit in \Cref{fig:circuit-example}(a) uses three qubits and applies three gates: the \emph{Hadamard} (\coqe{H}) gate and two \emph{controlled-not} (\coqe{CNOT}) gates. The semantics of a gate is a \emph{unitary matrix} (a matrix that preserves the unitarity invariant of quantum states); applying a gate to a state is tantamount to multiplying the state vector by the gate's matrix. The matrix corresponding to the circuit in \Cref{fig:circuit-example}(a) is shown in \Cref{fig:circuit-example}(g), where $I$ is the $2 \times 2$ identity matrix, $CNOT$ is the matrix corresponding to the \coqe{CNOT} gate, and $H$ is the matrix corresponding to the \coqe{H} gate. 

A special, non-unitary \emph{measurement} operation is used to extract classical information from a quantum state (often, when a computation completes). Measurement collapses the state to one of the basis states with a probability related to the state's amplitudes. For example, measuring $\frac{1}{\sqrt{2}}(\ket{0} + \ket{1})$ will collapse the state to $\ket{0}$ with probability $\frac{1}{2}$ and likewise for $\ket{1}$, returning classical values 0 or 1, respectively.
The semantics of a program involving measurement amounts to a probability distribution over quantum states; such a distribution is called a \emph{mixed state}. In our example above, measurement produces a mixed state that is a uniform distribution over $\ket{0}$ and $\ket{1}$. By contrast, \emph{pure states} like $\ket{0}$ and $\frac{1}{\sqrt{2}}(\ket{0} + \ket{1})$ can be produced without measurement. \Cref{sec:measurement} discusses non-unitary semantics further.

\section{SQIR: A Small Quantum Intermediate Representation}
\label{sec:sqir}

\sqir is a simple quantum language deeply embedded in the Coq proof assistant. This section presents \sqir's syntax and semantics. We defer a detailed discussion of \sqir's design rationale to the next section. 

\subsection{Unitary SQIR: Syntax}
\label{sec:syntax}

\sqir's \emph{unitary fragment} is a sub-language of full \sqir for expressing programs consisting of unitary gates. (The full \sqir language extends unitary \sqir with measurement.)
A program in the unitary fragment has type \coqe{ucom} (for ``unitary command''), which we define in Coq as follows:
\begin{coq}
Inductive ucom (U: nat -> Set) (d : nat) : Set :=
| useq  : ucom U d -> ucom U d -> ucom U d
| uapp1 : U 1 -> nat -> ucom U d
| uapp2 : U 2 -> nat -> nat -> ucom U d
\end{coq}
The \coqe{useq} constructor sequences two commands; we use notational shorthand \coqe{p1 ; p2} for \coqe{useq p1 p2}. The two \coqe{uappn} constructors indicate the application of a quantum gate to $n$ qubits, where $n$ is 1 or 2. Qubits are identified as numbered indices into a \emph{global qubit register} of size $d$, which stores the quantum state. 
Gates are drawn from parameter \coqe{U}, which is indexed by a gate's size. For writing and verifying programs, we use the following \coqe{base} set for \coqe{U}, inspired by IBM's OpenQASM \cite{Cross2017}:\footnote{It is helpful for proofs to keep \coqe{U} small because the number of cases in the proof about a value of type \coqe{ucom U d} will depend on the number of gates in \coqe{U}. In our work on \voqc \cite{VOQC}, we define optimizations over a larger gate set that includes common gates like Hadamard, but convert these gates to our \coqe{base} set for proof.}
\begin{coq}
Inductive base : nat -> Set := 
  | U_R ($\theta ~ \phi ~ \lambda$ : R) : base 1 
  | U_CNOT          : base 2.
\end{coq}
That is, we have a one-qubit gate \coqe{U_R} (which we write $U_R$ when using math notation), which takes three real-valued arguments, and the standard two-qubit \emph{controlled-not} gate, \coqe{U_CNOT} (written $CNOT$ in math notation), which negates the second qubit wherever the first qubit is $\ket{1}$, making it the quantum equivalent of a \emph{xor} gate. The \coqe{U_R} gate can be used to express any single-qubit gate (see \Cref{sec:unitarysem}). 
Together, \coqe{U_R} and \coqe{U_CNOT} form a \emph{universal} gate set, meaning that they can be composed to describe any unitary operation~\cite{barenco1995}.

\paragraph*{Example: SWAP}
The following Coq function produces a unitary \sqir program that applies three controlled-not gates in a row, with the effect of exchanging two qubits in the register. We define \coqe{CNOT} as shorthand for \coqe{uapp2 U_CNOT}.
\begin{coq}
Definition SWAP d a b : ucom base d := CNOT a b; CNOT b a; CNOT a b.
\end{coq}

\paragraph*{Example: GHZ}

\Cref{fig:circuit-example}(b) is the \sqir representation of the circuit in \Cref{fig:circuit-example}(a), which prepares the three-qubit GHZ state~\cite{Greenberger1989}.
We describe \emph{families} of \sqir circuits by meta-programming in the Coq host language. 
The Coq function in \Cref{fig:circuit-example}(c) produces a \sqir program that prepares the $n$-qubit GHZ state, producing the program in \Cref{fig:circuit-example}(b) when given input 3.
In Figures~1(b--c), \coqe{H} and \coqe{I} apply the \coqe{U_R} encodings of the Hadamard and identity gates.


\subsection{Unitary SQIR: Semantics}
\label{sec:unitarysem}

Each $k$-qubit quantum gate corresponds to a $2^k \times 2^k$ unitary matrix. The matrices for our \coqe{base} set are:
\[ \denote{U_R(\theta, \phi, \lambda)} = \begin{pmatrix} 
    \cos(\theta/2) & -e^{i\lambda}\sin(\theta/2) \\
    e^{i\phi}\sin(\theta/2) & e^{i(\phi+\lambda)}\cos(\theta/2) 
  \end{pmatrix}, \qquad \denote{CNOT} = 
\begin{pmatrix} 
1 & 0 & 0 & 0 \\ 
0 & 1 & 0 & 0 \\ 
0 & 0 & 0 & 1 \\ 
0 & 0 & 1 & 0 
\end{pmatrix}.
\]

Conveniently, the $U_R$ gate can encode any single-qubit gate~\cite[Chapter 4]{NCbook}. For instance, two commonly-used single-qubit gates are $X$ (``not'') and $H$ (``Hadamard''). The former has the matrix $\begin{psmallmatrix} 0 & 1 \\ 1 & 0 \end{psmallmatrix}$ and serves to flip a qubit's $\alpha$ and $\beta$ amplitudes; it can be encoded as $U_R(\pi, 0, \pi)$. The $H$ gate has the matrix $\frac{1}{\sqrt{2}}\begin{psmallmatrix} 1 & 1 \\ 1 & -1 \end{psmallmatrix}$, and is often used to put a qubit into superposition (it takes $\ket{0}$ to $\frac{1}{\sqrt{2}}(\ket{0} + \ket{1})$); it can be encoded as $U_R(\pi/2,0,\pi)$. Multi-qubit gates are easily produced by combinations of $CNOT$ and $U_R$; we show the definition of the three-qubit ``Toffoli'' gate in \Cref{sec:vecstates}. 
Keeping our gate set small simplifies the language and enables easy case analysis---and does not complicate proofs.
We rarely unfold the definition of gates like $X$ or the three-qubit Toffoli, instead providing automation to directly translate these gates to their intended denotations. Hence, $X$ is translated directly to $\begin{psmallmatrix} 0 & 1 \\ 1 & 0 \end{psmallmatrix}$. Users can thereby easily extend \sqir with new gates and denotations.

A unitary \sqir program operating on a size-$d$ register corresponds to a $2^d \times 2^d$ unitary matrix.  Function \coqe{uc_eval} denotes the matrix corresponding to program \coqe{c}.
\begin{coq}
Fixpoint uc_eval {d} (c : ucom base d) : Matrix (2^d) (2^d) := ...
\end{coq}
We write $\denote{\mathtt{c}}_d$ for \coqe{uc_eval d c}. 
The denotation of composition is simple matrix multiplication:
\coqe{[[U1; U2]]d} $=$ \coqe{[[U2]]d} $\times$ \coqe{[[U1]]d}.
The denotation of \coqe{uapp1} is the denotation of its argument gate, but padded with the identity matrix so it has size $2^d \times 2^d$. To be precise, we have:
\begin{align*}
    \denote{\code{uapp1 U q}}_d&=\begin{cases}
                          I_{2^q} \otimes \denote{U} \otimes I_{2^{d-q-1}} & q < d \\
                          0_{2^{d}} &\text{otherwise}
                          \end{cases}
\end{align*}
where $I_n$ is the $n \times n$ identity matrix.
In the case of our \coqe{base} gate set, $\denote{U}$ is the $U_R$ matrix shown above.
The denotation of any gate applied to an out-of-bounds qubit is the zero matrix, ensuring that a circuit corresponds to a zero matrix if and only if it is ill-formed. We likewise prove that every well-formed circuit corresponds to a unitary matrix.

As our only two-qubit gate in the \coqe{base} set is \coqe{U_CNOT}, we specialize our semantics for \coqe{uapp2} to this gate.
To compute \coqe{[[CNOT q1 q2]]d}, we first decompose the $CNOT$ matrix into $\begin{psmallmatrix} 1 & 0 \\ 0 & 0\end{psmallmatrix}\otimes I_2 + \begin{psmallmatrix} 0 & 0 \\ 0 & 1 \end{psmallmatrix}\otimes X$.
We then pad the expression appropriately, obtaining the following when $q_1 < q_2 < d$:
\[
I_{2^{q_1}} \otimes \begin{psmallmatrix} 1 & 0 \\ 0 & 0\end{psmallmatrix} \otimes I_{2^{q_2 - q_1 - 1}} \otimes I_2 \otimes I_{2^{d - q_2 - 1}} ~+~
I_{2^{q_1}} \otimes \begin{psmallmatrix} 0 & 0 \\ 0 & 1 \end{psmallmatrix} \otimes I_{2^{q_2 - q_1 - 1}} \otimes X \otimes I_{2^{d - q_2 - 1}}.
\]
When $q_2 < q_1 < d$, we obtain a symmetric expression, and 
when 
either qubit is out of bounds, we get the zero matrix. 
Additionally, since the two inputs to $CNOT$ cannot be the same, if $q_1 = q_2$ we also obtain the zero matrix.

\paragraph*{Example: Verifying SWAP}
We can prove in Coq that \coqe{SWAP 2 0 1}, which swaps the first and second qubits in a two-qubit register, behaves as expected on two unentangled qubits:
\begin{coq}
Lemma swap2: forall (φ ψ : Vector 2), WF_Matrix φ -> WF_Matrix ψ -> 
  [[SWAP 2 0 1]]$_2$ × (φ ⊗ ψ) = ψ ⊗ φ.
\end{coq}
\coqe{WF_Matrix} says that $\phi$ and $\psi$ are well-formed vectors \cite[Section 2]{Rand2018a}. 
This proof can be completed by simple matrix multiplication. 
In the full development we prove the correctness of \coqe{SWAP d a b} for arbitrary dimension \coqe{d} and qubits \coqe{a} and \coqe{b}.

\subsection{Full SQIR: Adding Measurement}
\label{sec:measurement}

The full \sqir language adds a branching measurement construct inspired by Selinger's QPL~\cite{Selinger2004}.
This construct permits measuring a qubit, taking one of two branches based on the measurement outcome. 
Full \sqir defines ``commands'' \coqe{com} as either a unitary sub-program, a no-op \coqe{skip}, branching measurement, or a sequence of these.
\begin{coq}
Inductive com (U: nat -> Set) (d : nat) : Set :=
| uc   : ucom U d -> com U d
| skip : com U d
| meas : nat -> com U d -> com U d -> com U d
| seq  : com U d -> com U d -> com U d.
\end{coq}
The command \coqe{meas q $P_1$ $P_2$} measures qubit \coqe{q} and performs $P_1$ if the outcome is 1 and $P_2$ if it is 0.
We define non-branching measurement and resetting to a zero state in terms of branching measurement:
\begin{coq}
Definition measure q := meas q skip skip.
Definition reset q := meas q (X q) skip.
\end{coq}
As before, we use our \coqe{base} set of unitary gates for full \sqir.

\paragraph*{Example: Flipping a Coin}

It is simple to generate a random coin flip with a quantum computer: Use the Hadamard gate to put a qubit into equal superposition $\frac{1}{\sqrt{2}}(\ket{0} + \ket{1})$ and then measure it.
\begin{coq}
Definition coin : com base 1 := H 0; measure 0.
\end{coq}

\paragraph*{Density Matrix Semantics}

As discussed in \Cref{sec:qprograms}, measurement induces a probabilistic transition, so the semantics of a program with measurement is a probability distribution over states, called a mixed state. As is standard~\cite{Paykin2017, Ying2011}, we represent such a state using a \emph{density matrix}. The density matrix of a pure state $\ket{\psi}$ is $\ketbra{\psi}{\psi}$ where $\bra{\psi} = \ket{\psi}^\dagger$ is the conjugate transpose of $\ket{\psi}$. The density matrix of a mixed state is a sum over its constituent pure states. For example, the density matrix corresponding to the uniform distribution over $\ket{0}$ and $\ket{1}$ is $\frac{1}{2}\ketbra{0}{0} + \frac{1}{2}\ketbra{1}{1}$. 

The semantics \coqe{[[[P]]]}$_d$ of a full \sqir program \coqe{P} is a function from density matrices to density matrices. Naturally, \coqe{[[[skip]]]d ρ = ρ} and \coqe{[[[P1 ; P2]]]d = [[[P2]]]d ∘ [[[P1]]]d}. For unitary subroutines, we have \coqe{[[[uc U]]]d ρ  = [[U]]dρ[[U]]d†}: Applying a unitary matrix to a state vector is equivalent to applying it to both sides of its density matrix.
Finally, using $\ket{i}_q\!\bra{j}$ for $I_{2^q} \otimes \ketbra{i}{j} \otimes I_{2^{d-q-1}}$, the semantics for \coqe{[[[meas q P1 P2]]]d ρ} is
\[
\pdenote{P_1}_d(\ket{1}_q\!\bra{1}\rho\ket{1}_q\!\bra{1}) + \pdenote{P_2}_d(\ket{0}_q\!\bra{0}\rho\ket{0}_q\!\bra{0})
\]
which corresponds to probabilistically applying \coqe{P1} to $\rho$ with the specified qubit projected to $\ketbra{1}{1}$ or applying \coqe{P2} to a similarly altered $\rho$. 

\paragraph*{Example: A Provably Random Coin}
We can now prove that our \coqe{coin} circuit above produces the $\ketbra{1}{1}$ or $\ketbra{0}{0}$ density matrix (corresponding to the $\ket{1}$ or $\ket{0}$ pure state), each with probability $\frac{1}{2}$.
\begin{coq}
Lemma coin_dist : [[[coin]]]$_1$ \k0\b0 = $\frac{1}{2}$\k1\b1 + $\frac{1}{2}$\k0\b0.
\end{coq}
The proof proceeds by simple matrix arithmetic. \coqe{[[[H]]] \k0\b0} is $H\ketbra{0}{0}H^\dagger = \frac{1}{2}\begin{psmallmatrix}1 & 1 \\ 1 & 1\end{psmallmatrix}$. Calling this $\rho_{12}$, applying \coqe{measure} yields $\ketbra{1}{1}\rho_{12}\ketbra{1}{1} + \ketbra{0}{0}\rho_{12}\ketbra{0}{0}$, which can be
further simplified using the fact $\bra{1}\rho_{12}\ket{1} = \bra{0}\rho_{12}\ket{0} = \begin{psmallmatrix} \frac{1}{2} \end{psmallmatrix}$, yielding $\frac{1}{2}\ketbra{1}{1} + \frac{1}{2}\ketbra{0}{0}$ as desired.

Measurement plays a key role in many quantum algorithms; we discuss further examples and an alternative semantics in \aref{app:measurement}.

\section{SQIR's Design}
\label{sec:design}

This section describes key elements in the design of \sqir and its infrastructure for verifying quantum programs. To place those decisions in context, we first introduce several related verification frameworks and contrast \sqir's design with theirs. 
In summary, \sqir benefits from the use of \emph{concrete indices into a global register} (a common feature in the tools we looked at), support for \emph{reasoning about unitary programs in isolation} (supported by one other tool), and the \emph{flexibility to allow different semantics and approaches to proof} (best supported in \sqir).

\subsection{Related Approaches}
\label{sec:related}

Several prior works have had the goal of formally verifying quantum programs. 
In 2010, Green~\cite{Green2010} developed an Agda implementation of the Quantum IO Monad, and in 2015 Boender et al.~\cite{Boender2015} produced a small Coq quantum library for reasoning about quantum ``programs'' directly via their matrix semantics (e.g. see \Cref{fig:circuit-example}(g)). These were both proofs of concept, and were only capable of verifying basic protocols.
More recently, Bordg et al.~\cite{bordg2020} took a step further in verifying quantum programs expressed as matrix products (\Cref{fig:circuit-example}(g)), providing a library for reasoning about quantum computation in Isabelle/HOL and verifying more interesting protocols like the $n$-qubit Deutsch-Jozsa algorithm (shown in \sqir in \aref{app:general-verification}).

In this section, we compare \sqir's design against three other tools for verified quantum programming that have been used to verify interesting, parameterized quantum programs:  \qwire~\cite{RandThesis} (implemented in Coq~\cite{coq}); quantum Hoare logic \cite{Liu2019AFP} (in Isabelle/HOL~\cite{isabelle}); and \qbricks~\cite{qbricks} (in Why3~\cite{Why3}).
We do not include Bordg et al.~\cite{bordg2020}, despite its recency, because it operates one level below the surface programming language, so many issues considered here do not apply.
Bordg et al.'s library is similar to the quantum libraries developed for \qwire and the quantum Hoare logic. 
All matrix formalisms provided by Bordg et al. are also available in \qwire's library, which we re-use and extend (by $\sim 3000$ lines \cite[Section 2.2]{VOQC}) in \sqir.

\paragraph*{QWIRE}

The \qwire language~\cite{Paykin2017,RandThesis} originated as an embedded circuit description language in the style of Quipper~\cite{Green2013} but with a more powerful type system. \Cref{fig:circuit-example}(d) shows the \qwire equivalent of the \sqir program in \Cref{fig:circuit-example}(b). \qwire uses variables from the host language Coq to reference qubits, an instantiation of higher-order abstract syntax~\cite{Pfenning1988}. In \Cref{fig:circuit-example}, the \qwire program uses variables \coqe{x}, \coqe{y}, and \coqe{z}, while the \sqir program uses indices \coqe{0}, \coqe{1}, and \coqe{2} to refer to the first, second, and third qubits in the global register. \qwire does not distinguish between unitary and non-unitary programs, and thus uses density matrices for its semantics. \qwire has been used to verify simple randomness generation circuits and a few textbook examples~\cite{Rand2017}. 

\paragraph*{QBRICKS}

\qbricks~\cite{qbricks} is a quantum proof framework implemented in Why3~\cite{Why3}, developed concurrently with \sqir. 
\qbricks provides a domain-specific language (DSL) for constructing quantum circuits using combinators for parallel and sequential composition (among others).
\Cref{fig:circuit-example}(e) presents the GHZ example written in \qbricks' DSL\@.
The semantics of \qbricks are based on the \emph{path-sums} formalism by Amy \cite{AmyThesis,Amy2018a}, which can express the semantics of unitary programs in a form amenable to proof automation. \qbricks extends path-sums to support parameterized circuits. \qbricks has been used to verify a variety of quantum algorithms, including Grover's algorithm and Quantum Phase Estimation~(QPE).

\paragraph*{Quantum Hoare Logic}

Quantum Hoare logic (QHL) \cite{Ying2011} has been formalized in the Isabelle/HOL proof assistant~\cite{Liu2019}. 
QHL is built on top of the quantum while language (QWhile), which is the quantum analog of the classical while language, allowing looping and branching on measurement results. \Cref{fig:circuit-example}(f) presents the GHZ example written in QHL. 
QWhile does not use a fixed gate set; gates are instead described directly by their unitary matrices. As such, the program in \Cref{fig:circuit-example}(f) could instead be written as the application of a single gate that prepares the 3-qubit GHZ state.
Given that measurement is a core part of the language, QWhile's semantics are given in terms of (partial) density matrices. A density matrix is \emph{partial} when it may represent a sub-distribution---that is, a subset of the outcomes of measurement.

QHL has been used to verify Grover's algorithm~\cite{Liu2019}. An earlier effort by Liu et al. \cite{Liu2016} to formalize QHL claimed to prove correctness of QPE, too. However, the approach used a combination of Isabelle/HOL and Python, calling out to Numpy to solve matrix (in)equalities; as such, we consider this only a partial verification effort. We cannot find a proof of QPE in the associated Github repository\footnote{\url{https://github.com/ijcar2016/propitious-barnacle}} and believe that this approach was abandoned in favor of Liu et al. \cite{Liu2019}.

\subsection{Concrete Indices into a Global Register}
\label{sec:concrete-indices}


The first key element of \sqir's design is its use of concrete indices into a fixed-sized global register to refer to qubits. For example, in our \coqe{SWAP} program (end of \Cref{sec:syntax}), \coqe{a} and \coqe{b} are natural numbers indexing into a global register of size \coqe{d}. 
Expressing the semantics of a program that uses concrete indices is simple because concrete indices map directly to the appropriate rows and columns in the denoted matrix.  Moreover, it is easy to check relationships between operations---\coqe{X a} and \coqe{X b} act on the same qubit if and only if \coqe{a $=$ b}. Keeping the register size fixed means that the denoted matrix's size is known, too.

On the other hand, concrete indices hamper programmability. The \coqe{ghz} example in \Cref{fig:circuit-example}(c) only produces circuits that occupy global qubits $0 ... $\coqe{n}; we could imagine further generalizing it to add a lower bound \coqe{m} (so the circuit uses qubits \coqe{m} $...$ \coqe{n}), but it is not clear how it could be generalized to use non-contiguous wires. A natural solution, employed by \qwire, is to use host-level variables to refer to \emph{abstract} qubits that can be freely introduced and discarded, simplifying circuit construction and sub-program composition.  Unfortunately, abstract qubits significantly complicate formal verification. To translate circuits to operations on density matrices, variables must be mapped to concrete matrix indices. Each time a qubit is discarded, indices undergo a de Bruijn-style shifting.

Similar to \sqir's use of concrete indices, \qbricks-DSL's compositional structure makes it easy to map programs to their denotation: The ``index'' of a gate application can be computed by its nested position in the program. However, this syntax is even less convenient than \sqir's for programming: Although \qbricks provides a utility function for defining \coqe{CNOT} gates between non-adjacent qubits, their underlying syntax does not support this, meaning that expressions like \coqe{CNOT 7 2} are translated into large sequences of \coqe{CNOT} gates. QHL is presented as having variables (e.g. \coqe{q1} in \Cref{fig:circuit-example}(f)), but these variables are fixed before a program is executed and persist throughout the program. In the Isabelle formalization, they are represented by natural numbers, making them comparable to \sqir concrete indices.

\subsection{Extensible Language around a Unitary Core}
\label{sec:unitarycore}

Another key aspect of \sqir's design is its decomposition into a unitary sub-language and the non-unitary full language. While the full language (with measurement) is more powerful, its density matrix-based semantics adds unneeded complication to the proof of unitary programs. For example, given the program $U_1; U_2; U_3$, its unitary semantics is a matrix $U_3 \times U_2 \times U_1$ while its density matrix semantics is a function $\rho \mapsto U_3 \times U_2 \times U_1 \times \rho \times U_1^\dagger \times U_2^\dagger \times U_3^\dagger$. The latter is a larger term, with a type that is harder to work with. This added complexity, borne by \qwire and QHL, lacks a compelling justification given that many algorithms can be viewed as unitary programs with measurement occurring implicitly at their conclusion (see \Cref{sec:meas-predicates}).

On the other hand, \qbricks' semantics is based on (higher-order) path-sums, which cannot describe mixed states, and thus cannot give a semantics to measurement. \sqir's design allows for a ``best of both worlds,'' utilizing a unitary semantics when possible, but supporting non-unitary semantics when needed. Furthermore, as we show in \cref{sec:vecstates}, abstractions like path-sums can be easily defined on top of \sqir's unitary semantics.

\subsection{Semantics of Ill-typed Programs}
\label{sec:ill-typed}

We say that a \sqir program is well-typed if every gate is applied to indices within range of the global register and indices used in each multi-qubit gate are distinct. 
This second condition enforces quantum mechanics' \emph{no-cloning theorem}, which disallows copying an arbitrary quantum state,
as would be required to evaluate an expression like \coqe{CNOT q q}.
For example, \coqe{SWAP d a b} is well-typed if $a < d$, $b < d$, and $a \neq b$.

\qwire addresses this issue through its linear type system, which also guarantees that qubits are never reused. However, well-typedness is a (non-trivial) 
extrinsic proposition 
in \qwire, meaning that many proofs require an 
assumption that the input program is well-typed and must manipulate this typing judgment within the proof. 
\qbricks avoids the issue of well-typedness through its language design: It is not possible to construct an ill-typed circuit using sequential and parallel composition. 
The Isabelle implementation of QHL uses a well-typedness predicate to enforce some program restrictions (e.g. the gate in a unitary application is indeed a unitary matrix), but the issue of gate argument validity is enforced by Isabelle's type system: Gate arguments are represented as a set (disallowing duplicates) where all elements are valid variables. 

In \sqir, ill-typed programs are denoted by the zero matrix.
This often means that we do not need to explicitly assume or prove that a program is well-typed in order to state a property about its semantics, thereby removing clutter from theorems and proofs.
For example, we can prove symmetry of \coqe{SWAP}, i.e. \coqe{SWAP d a b $\equiv$ SWAP d b a}, without any well-typedness constraint because either both sides of the equation are well-typed or both are ill-typed.
However, we cannot always avoid well-typedness preconditions. Say we want to prove transitivity of \mbox{\coqe{SWAP},} i.e. \coqe{SWAP d a c $\equiv$ SWAP d a b ; SWAP d b c}.
In this case the left-hand side may be well-typed while the right-hand side is ill-typed. To verify this equivalence, we (minimally) need the precondition \coqe{b < d /\ b <> a /\ b <> c}. 
We capture these in our \coqe{uc_well_typed} predicate, which resembles the \coqe{WF_Matrix} predicate (used in the \coqe{SWAP} example in \Cref{sec:unitarysem}) that guarantees that a matrix's non-zero entries are all within its bounds~\cite[Section 3.3]{VOQC}. Both conditions are easily checked via automation.  

\subsection{Automation for Matrix Expressions}
\label{sec:automation}

The \sqir development provides a variety of automation techniques for dealing with matrix expressions.
Most of this automation is focused on simplifying matrix terms to be easier to work with.
The best example of this is our \coqe{gridify} tactic \cite[Section 4.5]{VOQC}, which rewrites terms into \emph{grid normal form} where matrix addition is on the outside, followed by tensor product, with matrix multiplication on the inside, i.e., $((..\times..)\otimes(..\times..)) + ((..\times..)\otimes(..\times..))$. 
Most of the circuit equivalences available in \sqir (e.g. $\forall$ \coqe{a, b, c}. \coqe{CNOT a c ; CNOT b c $\equiv$ CNOT b c ; CNOT a c}) are proved using \coqe{gridify}.
This style of automation is available in other verification tools too; \coqe{gridify} is similar to Liu et al.'s Isabelle tactic for matrix normalization \cite[Section 5.1]{Liu2019}.
\qbricks avoids the issue by using path-sums; they provide a matrix semantics for comparison's sake, but do not discuss automation for it.

Some of our automation is aimed at alleviating difficulties caused by our use of \emph{phantom types}~\cite{Rand2018a} to store the dimensions of a matrix, the rationale of which is explained in our prior work \cite[Section 3.3]{VOQC}.
In our development, matrices have the type \coqe{Matrix m n}, where \coqe{m} is the number of rows and \coqe{n} is the number of columns. 
One challenge with this definition is that the dimensions stored in the type may be ``out of sync'' with the structure of the expression itself.
For example, due to simplification, rewriting, or declaration, the expression $\ket{0} \otimes \ket{0}$ may be annotated with the type \coqe{Vector 4}, although rewrite rules expect it to be of the form \coqe{Vector (2 * 2)}.
We provide a tactic \coqe{restore_dims} that analyzes the structure of a term and rewrites its type to the desired form, allowing for more effective automated simplification. 

\subsection{Vector State Abstractions}
\label{sec:vecstates}

To verify that the \coqe{SWAP} program has the intended semantics, we can unfold its definition (\coqe{CNOT a b; CNOT b a; CNOT a b}) and compute the associated matrix expression.
However, while this proof is made simpler by automation like \coqe{gridify}, it is still fairly complicated considering that \coqe{SWAP} has a simple classical (non-quantum) purpose.
In fact, this operation is much more naturally analyzed using its action on basis states.
A \emph{(computational) basis state} is any state of the form $\ket{i_1 \dots i_d}$ for $i_1, \dots, i_d \in \{0, 1\}$ (so $\ket{00}$ and $\ket{11}$ are basis states, while $\frac{1}{\sqrt{2}}(\ket{00} + \ket{11})$ is not).
The set of all $d$-qubit basis states form a basis for the underlying $d$-dimensional vector space, meaning that any $2 ^ d \times 2 ^ d$ unitary operation can be uniquely described by its action on those basis states.

Using basis states, the reasoning for our \coqe{SWAP} example proceeds as follows, where we use $\ket{\dots x \dots y \dots}$ as informal notation to describe the state where the qubit at index $a$ is in state $x$ and the qubit at index $b$ is in state $y$.
\begin{enumerate}
    \item Begin with the state $\ket{\dots x \dots y \dots}$.
    \item $CNOT~ a~ b$ produces $\ket{\dots x \dots (x \oplus y) \dots}$.
    \item $CNOT~ b~ a$ produces $\ket{\dots (x \oplus (x \oplus y)) \dots (x \oplus y) \dots} = \ket{\dots y \dots (x \oplus y) \dots}$.
    \item $CNOT~ a~ b$ produces $\ket{\dots y \dots (y \oplus (x \oplus y)) \dots} = \ket{\dots y \dots x \dots}$.
\end{enumerate}
In our development, we describe basis states using \coqe{f_to_vec d f} where \coqe{d : nat} and \coqe{f : nat -> bool}. 
This describes a $d$-qubit quantum state where qubit $i$ is in the basis state $f(i)$, and \coqe{false} corresponds to $0$ and \coqe{true} to $1$. 
We also sometimes describe basis states using \coqe{basis_vector d i} where 
$i < 2^d$ is the index of the only $1$ in the vector. 
We provide methods to translate between the two representations (simply converting between binary and decimal encodings).
For the remainder of the paper, we will write \coqe{$\ket{\text{f}}$} for \coqe{f_to_vec n f} and \coqe{$\ket{\text{i}}$} for \coqe{basis_vector n i}, omitting the \coqe{n} parameter when it is clear from the context.

We prove a variety of facts about the actions of gates on basis states. For example, the following succinctly describe the behavior of the $CNOT$ and $Rz(\theta)$ gates, where $Rz(\theta) = U_R(0,0,\theta)$:
\begin{coq}
Lemma f_to_vec_CNOT : forall (d i j : nat) (f : nat -> bool),
  i < d -> j < d -> i <> j ->
  let f' := update f j (f j ⊕ f i) in
  [[CNOT i j]]$_d$ × $\ket{\text{f}}$ = $\ket{\text{f\textquotesingle}}$.
      
Lemma f_to_vec_Rz: forall (d j : nat) (θ : R) (f : nat -> bool),
  j < d ->
  [[Rz θ j]]$_d$ × $\ket{\text{f}}$ = $e ^ {i \theta (f~ j)}$ * $\ket{\text{f}}$.
\end{coq}

Above, \coqe{update f i v} updates the value of \coqe{f} at index \coqe{i} to be \coqe{v} (i.e. for the resulting function $f'$, $f'(i) = v$ and $f'(j) = f(j)$ for all $j \neq i$).
So $CNOT~i~j$ has the effect of updating the $j^\text{th}$ entry of the input state to the exclusive-or of its $i^\text{th}$ and $j^\text{th}$ entries.
$Rz~ \theta~ j$ updates the \emph{phase} associated with the input state.

There are several advantages to applying these rewrite rules instead of unfolding the definitions of \coqe{[[CNOT i j]]$_d$} and \coqe{[[Rz θ j]]$_d$}.
For example, these rewrite rules assume well-typedness and do not depend on the ordering of qubit arguments, avoiding the case analysis needed in \coqe{gridify} \cite[Section 4.5]{VOQC}.
In addition, the rule for $CNOT$ above is simpler to work with than the general unitary semantics ($CNOT \mapsto \_ \otimes \begin{psmallmatrix} 1 & 0 \\ 0 & 0\end{psmallmatrix} \otimes \_ \otimes I_2 \otimes \_ ~+~ \_ \otimes \begin{psmallmatrix} 0 & 0 \\ 0 & 1\end{psmallmatrix} \otimes \_ \otimes \sigma_x \otimes \_$).

As a concrete example of where vector-based reasoning was critical, consider the three-qubit Toffoli gate, which implements a \emph{controlled-controlled-not}, and can be thought of as the quantum equivalent of an \emph{and} gate. It is frequently used in algorithms, but (like all $n$-qubit gates with $n > 2$) rarely supported in hardware, meaning that it must be decomposed into more basic gates before execution. In practice, we found \coqe{gridify} too inefficient to verify the standard decomposition of the gate~\cite[Chapter 4]{NCbook}, shown below.
\begin{coq}
Definition TOFF {d} a b c : ucom base d :=
  H c ; CNOT b c ; T$^\dagger$ c ; CNOT a c ; T c ; CNOT b c ; T$^\dagger$ c ; 
  CNOT a c ; CNOT a b ; T$^\dagger$ b ; CNOT a b ; T a ; T b ; T c ; H c. 
\end{coq}
However, like \coqe{SWAP}, the semantics of the Toffoli gate is naturally expressed through its action on basis states:
\begin{coq}
Lemma f_to_vec_TOFF : forall (d a b c : nat) (f : nat -> bool),
  a < d -> b < d -> c < d -> 
  a <> b -> a <> c -> b <> c ->
  let f' := update f c (f c ⊕ (f a && f b)) in
  [[TOFF a b c]]$_d$ × $\ket{\text{f}}$ = $\ket{\text{f\textquotesingle}}$.
\end{coq}
The proof of \coqe{f_to_vec_TOFF} is almost entirely automated using a tactic that rewrites using the \coqe{f_to_vec} lemmas shown above, since \coqe{T} and \coqe{T$^\dagger$} are \coqe{Rz (PI / 4)} and \coqe{Rz (-PI / 4)}, respectively.

The \coqe{f_to_vec} abstraction is simple and easy to use, but not universally applicable: Not all quantum algorithms produce basis states, or even sums over a small number of basis states, and reasoning about $2^d$ terms of the form $\ket{i_1 \dots i_d}$ is no easier than reasoning directly about matrices. 
To support more general types of quantum states we define indexed sums and tensor (Kronecker) products of vectors.
\begin{coq}
Fixpoint vsum {d} n (f: nat -> Vector d) : Vector d := ...
Fixpoint vkron n (f: nat -> Vector 2) : Vector $2^n$ := ...
\end{coq}

As an example of a state that uses these constructs, the action of $n$ parallel Hadamard gates on the state $\ket{f}$ can be written as
\begin{center}
\coqe{vkron n (fun i => $\frac{1}{\sqrt{2}}$($\ket{0}$ + $(-1)^{f(i)}\ket{1}$))}
~ or ~
\coqe{$\frac{1}{\sqrt{2^n}}$ * (vsum $2^n$ (fun i => $(-1)^{\code{to\_int}(f) \bullet i}$ * $\ket{i}$))},
\end{center}
both commonly-used facts in quantum algorithms.
For the remainder of the paper, we will write $\sum_{i=0}^{n-1} f(i)$ for \coqe{vsum n (fun i => f i)} and $\bigotimes_{i=0}^{n-1} f(i)$ for \coqe{vkron n (fun i => f i)}.

\paragraph*{Relation with Path-sums}

Our \coqe{vsum} and \coqe{vkron} definitions share similarities with the \emph{path-sums}~\cite{AmyThesis,Amy2018a} semantics used by \qbricks~\cite{qbricks}.
In the path-sums formalism, every unitary transformation is represented as a function of the form
\[
\ket{x} \to \frac{1}{\sqrt{2^m}} \sum_{y = 0}^{2^m - 1}  e^{2\pi i P(x,y) / 2^m} \ket{f(x,y)}
\]
where $m \in \N$, $P$ is an arithmetic function over $x$ and $y$, and $f$ is of the form $\ket{f_1(x,y)} \otimes \dots \otimes \ket{f_m(x,y)} $ where each $f_i$ is a Boolean function over $x$ and $y$. 
For instance, the Hadamard gate $H$ has the form $\ket{x} \to \frac{1}{\sqrt{2}} \sum_{y = 0}^{1}e^{2\pi i xy/2} \ket{y}$. 
Path-sums provide a compact way to describe the behavior of unitary matrices and are closed under matrix and tensor products, making them well-suited for automation.
They can be naturally described in terms of our \coqe{vkron} and \coqe{vsum} vector-state abstractions:
\begin{coq}
Definition path_sum (m : nat) P f x := 
  vsum $2^m$ (fun y => $e^{2\pi i P(x,y)/2^m}$ * (vkron m (fun i => f i x y))).
\end{coq}
As above, \coqe{P} is an arithmetic function over \coqe{x} and \coqe{y} and \coqe{f i} is a Boolean function over \coqe{x} and \coqe{y} for any \coqe{i}. 

\subsection{Measurement Predicates}
\label{sec:meas-predicates}

The proofs in \Cref{sec:examples} do not use the non-unitary semantics directly, but instead describe the probability of different measurement outcomes using predicates \coqe{probability_of_outcome} and \coqe{prob_partial_meas}.
\begin{coq}
(* Probability of measuring ϕ given input ψ. *)
Definition probability_of_outcome {n} (ϕ ψ : Vector n) : R :=
  let c := (ϕ† × ψ) 0 0 in $|c|^2$.
  
(* Probability of measuring ϕ on the first n qubits given (n+m) qubit input ψ. *)
Definition prob_partial_meas {n m} (ϕ : Vector $2^n$) (ψ : Vector $2^{n + m}$) :=
  $\lVert$ (ϕ† ⊗ I$_{2 ^ m}$) × ψ $\rVert^ 2$.
\end{coq}
Above, $\lVert v \rVert$ is the 2-norm of vector $v$ and $|c|$ is the complex norm of $c$.
In formal terms, the ``probability of measuring $\varphi$'' is the probability of outcome $\varphi$ when measuring a state in the basis $\{ \varphi \times \varphi^\dagger, \text{\coqe{I}}_{2^n}-\varphi\times \varphi^\dagger \}$. 

The \emph{principle of deferred measurement}~\cite[Chapter 4]{NCbook} says that measurement can always be deferred until the end of a quantum computation without changing the result.
However, we included measurement in \Cref{sec:measurement} because it is an important feature of quantum programming languages that is used in a variety of constructs like repeat-until-success loops~\cite{paetznick2014repeat} and error-correcting codes~\cite{gottesman2010introduction}. 
\qbricks also uses measurement predicates, but unlike \sqir does not support a general measurement construct.

\section{Proofs of Quantum Algorithms}
\label{sec:examples}

In this section we discuss the formal verification of two classic quantum algorithms: 
Grover's algorithm \cite[Chapter 6]{NCbook} and quantum phase estimation \cite[Chapter 5]{NCbook}. 
We present additional, simpler examples in \aref{app:measurement,app:general-verification}.
All proofs and specifications follow the corresponding textbook arguments.

\subsection{Grover's Algorithm}

\paragraph*{Overview}

Given a circuit implementing Boolean oracle $f : \{0,1\}^n\to\{0,1\}$, the goal of Grover's algorithm is to find an input $x$ satisfying $f(x) = 1$. Suppose that $n \geq 2$. In the classical (worst-)case where $f(x)=1$ has a unique solution, finding this solution requires $O(2^n)$ queries to the oracle. However, the quantum algorithm finds the solution with high probability using only $O(\sqrt{2^n})$ queries.

The algorithm alternates between applying the oracle and a ``diffusion operator.''
Individually, these operations each perform a reflection in the two-dimensional space spanned by the input vector (a uniform superposition) and a uniform superposition over the solutions to $f$. Together, they perform a rotation in the same space.
By choosing an appropriate number of iterations $i$, the algorithm will rotate the input state to be suitably close to the solution vector.
The \sqir definition of Grover's algorithm is shown in \Cref{fig:grov}.

\begin{figure}
    \centering
\begin{coq}
(* Controlled-X with target (n-1) and controls 0, 1, ..., n-2. *)
Fixpoint generalized_Toffoli' n0 : ucom base n :=
  match n0 with
  | O | S O => X (n - 1)
  | S n0' => control (n - n0) (generalized_Toffoli' n0')
  end.
Definition generalized_Toffoli := generalized_Toffoli' n.

(* Diffusion operator. *)
Definition diff : ucom base n :=
  npar n H; npar n X ; 
  H (n - 1) ; generalized_Toffoli ; H (n - 1) ; 
  npar n X; npar n H.

(* Main program (iterates applying Uf and diff). *)
Definition body := Uf ; cast diff (S n).
Definition grover i := X n ; npar (S n) H ; niter i body.
\end{coq}
    \caption{Grover's algorithm in \sqir. \coqe{control} performs a unitary program conditioned on an input qubit, \coqe{npar} performs copies of a unitary program in parallel, \coqe{cast} is a no-op that changes the dimension in a \coqe{ucom}'s type, and \coqe{niter} iterates a unitary program.}
    \label{fig:grov}
\end{figure}

The \sqir version of Grover's algorithm is 15 lines, excluding utility definitions like \coqe{control} and \coqe{npar}. 
The specification and proof are around 770 lines.
The proof took approximately one person-week.

\paragraph*{Proof Details}
The statement of correctness says that after $i$ iterations, the probability of measuring a solution is $\sin^2((2i + 1)\theta)$ where $\theta = \arcsin (\sqrt{k/2^n})$ and $k$ is the number of satisfying solutions to $f$. Note that this implies that the optimal number of iterations is $\frac{\pi}{4}\sqrt{\frac{2^n}{k}}$.

We begin the proof by showing that the uniform superposition can be rewritten as a sum of ``good'' states (\coqe{$\psi$g}) that satisfy $f$ and ``bad'' states (\coqe{$\psi$b}) that do not satisfy $f$.
\begin{coq}
Definition ψ := $\frac{1}{\sqrt{2 ^ n}}\sum_{k=0}^{2^n - 1}\ket{k}$.
Definition θ := asin ($\sqrt{k/2^n}$).
Lemma decompose_ψ : ψ = (sin θ) ψg + (cos θ) ψb.
\end{coq}
We then prove that \coqe{Uf} and \coqe{diff} perform the expected reflections (e.g. \coqe{[[diff]]$_n$} $= -2\ket{\psi}\bra{\psi} + I_{2^n}$) and the following key lemma, which shows the output state after $i$ iterations of \coqe{body}.
\begin{coq}
Lemma loop_body_action_on_unif_superpos : forall i,
  [[body]]$_{n + 1}^i$ (ψ ⊗ ∣-⟩) =
    (-1)$^i$ (sin ((2 * i + 1) * θ) ψg + cos ((2 * i + 1) * θ) ψb) ⊗ ∣-⟩.
\end{coq}
This property is straightforward to prove by induction on \coqe{i}, and implies the desired result, which specifies the probability of measuring any solution to $f$.

\begin{coq}
Lemma grover_correct : forall i,
  Rsum $2^n$ (fun z =>  if f z 
                     then prob_partial_meas $\ket{\text{z}}$ ([[grover i]]$_{n+1}$ × $\ket{\text{0}}^{n+1}$)
                     else 0) = 
  (sin ((2 * i + 1) * θ))$ ^ 2$.
\end{coq}
That is, the sum over the probability of all possible outcomes $z$ such that $f(z)$ is true is $\sin^2((2i + 1)\theta)$. Above, \coqe{Rsum} is a sum over real numbers.

\subsection{Quantum Phase Estimation}

\paragraph*{Overview}

Given a unitary matrix $U$ and eigenvector $\ket{\psi}$ such that $U\ket{\psi} = e^{2\pi i\theta}\ket{\psi}$, the goal of quantum phase estimation (QPE) is to find a $k$-bit representation of $\theta$. In the case where $\theta$ can be exactly represented using $k$ bits (i.e. $\theta = z / 2 ^ k$ for some $z \in \mathbb{Z}$), QPE recovers $\theta$ exactly. Otherwise, the algorithm finds a good $k$-bit approximation with high probability. QPE is often used as a subroutine in quantum algorithms, most famously Shor's factoring algorithm \cite{Shor94}. 

\begin{figure}[hp]
\centering
\begin{coq}
(* Controlled rotation cascade on n qubits. *)
Fixpoint controlled_rotations n : ucom base n :=
  match n with
  | 0 | 1 => SKIP
  | S n'  => controlled_rotations n' ; control n' (Rz ($2\pi$ / $2^n$) 0)
  end.

(* Quantum Fourier transform on n qubits. *)
Fixpoint QFT n : ucom base n :=
  match n with
  | 0    => SKIP
  | S n' => H 0 ; controlled_rotations n ; map_qubits (fun q => q + 1) (QFT n') 
  end.

(* The output of QFT needs to be reversed before further processing. *)
Definition reverse_qubits n := ...
Definition QFT_w_reverse n := QFT n ; reverse_qubits n.

(* Controlled powers of u. *)
Fixpoint controlled_powers' {n} (u : ucom base n) k kmax : ucom base (kmax+n) :=
  match k with
  | 0    => SKIP
  | S k' => controlled_powers' u k' kmax ; niter $2^{k'}$ (control (kmax - k' - 1) u)
  end.
Definition controlled_powers {n} (u : ucom base n) k := controlled_powers' u k k.

(* QPE circuit for program u.
   k = number of bits in resulting estimate
   n = number of qubits in input state *)
Definition QPE k n (u : ucom base n) : ucom base (k + n) :=
  npar k H ;
  controlled_powers (map_qubits (fun q => k + q) u) k; 
  invert (QFT_w_reverse k).
\end{coq}
\vspace{-0.5em}
\caption{\sqir definition of QPE. Some type annotations and calls to \coqe{cast} have been removed for clarity. \coqe{control}, \coqe{map_qubits}, \coqe{niter}, \coqe{npar}, and \coqe{invert} are Coq functions that transform \sqir programs; we have proved that they have the expected behavior (e.g. $\forall~ u. ~\denote{\text{invert}~u}_n = \denote{u}_n^\dagger$).}
\label{fig:full-qpe}

\begin{center}
\vspace{1em}
\begin{tabular}{c c}
  \begin{minipage}{0.15\linewidth}
  $QPE_{k,n} =$
  \end{minipage}%
  &
  \begin{minipage}{0.75\linewidth}
  \small
  \Qcircuit @C=0.7em @R=0.5em {
     \lstick{\ket{0}} & \gate{H} & \qw & \qw & \qw & \qw & \dots & & \ctrl{4} & \multigate{3}{QFT_k^{-1}} & \qw & \qw \\
     \lstick{\vdots} & & & & & & & & & & \vdots &  \\
     \lstick{\ket{0}} & \gate{H} & \qw & \qw & \ctrl{2} & \qw & \dots & & \qw & \ghost{QFT_k^{-1}} & \qw & \qw \\
     \lstick{\ket{0}} & \gate{H} & \qw & \ctrl{1} & \qw & \qw & \dots & & \qw & \ghost{QFT_k^{-1}} & \qw & \qw \\
     \lstick{\ket{\psi}} & {/^n} \qw & \qw & \gate{U^{2^0}} & \gate{U^{2^1}} & \qw & \dots & & \gate{U^{2^{k-1}}} & \qw & \qw & \qw \\
    }
  \end{minipage}%
\end{tabular} 

\vspace{1em}
\begin{tabular}{c c}
  \begin{minipage}{0.1\linewidth}
  $QFT_k =$
  
  \end{minipage}%
  &
  \begin{minipage}{0.8\linewidth}
  \small
  \Qcircuit @C=0.7em @R=0.1em {
     & \gate{H} & \gate{R_2} & \qw & \dots & & \gate{R_{k-1}} & \gate{R_{k}} & \qw & \qw & \qw & \qw & \qw & \qw & \qw & \qw & \qw & \qw & \qw & \qw & \qw \\
     & \qw & \ctrl{-1} & \qw & \dots & & \qw & \qw & \gate{H} & \qw & \dots & & \gate{R_{k-2}} & \gate{R_{k-1}} & \qw & \qw & \qw & \qw & \qw & \qw & \qw \\
     & \vdots & & & & & & & & & & & & & & & & & & & \\
     & \qw & \qw & \qw & \qw & \qw & \ctrl{-3} & \qw & \qw & \qw & \qw & \qw & \ctrl{-2} & \qw & \qw & \dots & & \gate{H} & \gate{R_2} & \qw & \qw \\
     & \qw & \qw & \qw & \qw & \qw & \qw & \ctrl{-4} & \qw & \qw & \qw & \qw & \qw & \ctrl{-3} & \qw & \dots & & \qw & \ctrl{-1} & \gate{H} & \qw\\
    }
  \end{minipage}%
\end{tabular}
\end{center}
  \vspace{-1em}
  \caption{Circuit for quantum phase estimation (QPE) with $k$ bits of precision and an $n$-qubit input state (top) and quantum Fourier transform (QFT) on $k$ qubits (bottom). $\ket{\psi}$ and $U$ are inputs to QPE. $R_m$ is a $z$-axis rotation by $2\pi / 2^m$.}
  \label{fig:qpe}

\end{figure}


The \sqir program for QPE is shown in \Cref{fig:full-qpe}. For comparison, the standard circuit diagrams for QPE and the quantum Fourier transform (QFT), which is used as a subroutine in QPE, are shown in \Cref{fig:qpe}. 
The \sqir version of QPE is around 40 lines and the specification and proof in the simple case ($\theta = z / 2 ^ k$) is around 800 lines. The fully general case ($\theta \neq z / 2 ^ k$) adds about 250 lines. The proof of the simple case was completed in about two person-weeks. 
When working out the proof of the general case, we found that we needed some non-trivial bounds on trigonometric functions (for $x \in \mathbb{R}$, $\abs{\sin(x)} \leq \abs{x}$ and if $\abs{x} \leq \frac{1}{2}$ then $\abs{2 * x} \leq \abs{\sin(\pi x )}$). 
Laurent Th\'{e}ry kindly provided proofs of these facts using the Coq Interval package \cite{intervals}.

\paragraph*{Proof Details}

The correctness property for QPE in the case where $\theta$ can be described exactly using $k$ bits ($\theta = z / 2^k$) says that the QPE program will exactly recover $z$.
It can be stated in \sqir's development as follows.
\begin{coq}
Lemma QPE_correct_simplified: forall k n (u : ucom base n) z (ψ : Vector $2^n$), 
  n > 0 -> k > 1 -> uc_well_typed u -> WF_Matrix ψ ->
  let θ := z / $2^k$ in
  [[u]]$_{n}$ × ψ = $e^{2\pi i\theta}$ * ψ ->
  [[QPE k n u]]$_{k+n}$ × (∣0⟩$^{k}$ ⊗ ψ) = $\ket{\text{z}}$ ⊗ ψ.
\end{coq}
The first four conditions ensure well-formedness of the inputs. The fifth condition enforces that input $\psi$ is an eigenvector of $c$. The conclusion says that running the \coqe{QPE} program computes the value $z$, as desired.

In the general case where $\theta$ cannot be exactly described using $k$ bits, we instead prove that \coqe{QPE} recovers the best $k$-bit approximation with high probability (in particular, with probability $\geq 4/\pi^2$). 
\begin{coq}
Lemma QPE_semantics_full : forall k n (u : ucom base n) z (ψ : Vector $2^n$) (δ : R),
  n > 0 -> k > 1 -> uc_well_typed u -> Pure_State_Vector ψ -> 
  -1 / $2^{k + 1}$ <= δ < 1 / $2^{k + 1}$ -> δ <> 0 ->
  let θ := z / $2^k$ + δ in
  [[u]]$_n$ × ψ = $e^{2\pi i\theta}$ * ψ ->
  prob_partial_meas $\ket{\text{z}}$ ([[QPE k n u]]$_{k+n}$ × (∣0⟩$^{k}$ ⊗ ψ)) >= 4 / $\pi^2$.
\end{coq}
\coqe{Pure_State_Vector} is a restricted form of \coqe{WF_Matrix} that requires a vector to have norm 1.

As an example of the reasoning that goes into proving these properties, consider the QFT subroutine of QPE.
The correctness property for \coqe{controlled_rotations} says that evaluating the program on input $\ket{x}$ will produce the state $e^{2\pi i (x_0 ~ \cdot ~  x_1x_2...x_{n-1})/2^{n}}\ket{x}$ where $x_0$ is the highest-order bit of $x$ represented as a binary string and $x_1x_2...x_{n-1}$ are the lower-order $n-1$ bits.
\begin{coq}
Lemma controlled_rotations_correct : forall n x,
  n > 1 -> [[controlled_rotations n]]$_n$ × ∣x⟩ = $e^{2\pi i (x_0 ~ \cdot ~  x_1x_2...x_{n-1})/2^{n}}$∣x⟩.
\end{coq}
We can prove this property via induction on $n$. In the base case ($n = 2$) we have that $x$ is a 2-bit string $x_0x_1$. In this case, the output of the program is $e ^ {2\pi i (x_0 \cdot x_1)/2^2}\ket{x_0x_1}$, as desired.
In the inductive step, we assume that: 
\begin{center}
    \coqe{[[controlled_rotations n]]$_n$ × ∣$x_1x_2...x_{n-1}$⟩ = $e^{2\pi i (x_0 ~ \cdot ~  x_1x_2...x_{n-1})/2^{n}}$∣$x_1x_2...x_{n-1}$⟩}.
\end{center}
We then perform the simplifications shown in \Cref{fig:reasoning}, which complete the proof.
\begin{figure*}
\centering
\begin{minipage}{\textwidth}
\begin{align*}
    &\denote{\code{controlled\_rotations (n+1)}}_{n+1} \times \ket{x} \\
    &\quad= \denote{\code{control}~ x_n~ (\code{Rz}~ (2\pi / 2^{n+1})~ 0}_{n+1} \times \denote{\code{controlled\_rotations n}}_{n+1} \times \ket{x} \\
    &\quad= \denote{\code{control}~ x_n~ (\code{Rz}~ (2\pi / 2^{n+1})~ 0}_{n+1} \times e^{2\pi i (x_0 ~ \cdot ~  x_1x_2...x_{n-1})/2^{n}}\ket{x_1x_2...x_{n-1}x_n} \\
    &\quad= e^{2\pi i (x_0 ~ \cdot ~  x_{n})/2^{n+1}}e^{2\pi i (x_0 ~ \cdot ~  x_1x_2...x_{n-1})/2^{n}}\ket{x_1x_2...x_{n-1}x_n} \\
    &\quad= e^{2\pi i (x_0 ~ \cdot ~  x_1x_2...x_n)/2^{n+1}}\ket{x_1x_2...x_{n-1}x_n}
\end{align*}
\end{minipage}
    \caption{Reasoning used in the proof of \coqe{controlled_rotations}. The first step unfolds the definition of \coqe{controlled_rotations}; the second step applies the inductive hypothesis; the third step evaluates the semantics of \coqe{control}; and the fourth step combines the exponential terms. }
    \label{fig:reasoning}
\end{figure*}

Our correctness property for \coqe{QFT n} (shown below) can similarly be proved by induction on $n$, and relies on the lemma \coqe{controlled_rotations_correct}.

\begin{coq}
Lemma QFT_semantics : forall n x, n>0 -> [[QFT n]]$_n$ × ∣x⟩ = $\frac{1}{\sqrt{2^n}} \bigotimes_{j=0}^{n-1} (\ket{\text{0}} + e^{2\pi i x / 2 ^ {n - j} } \ket{\text{1}}) $.
\end{coq}

\section{Open Problems and Future Work}
\label{sec:conclusions}

We previously presented \sqir as the intermediate representation in a verified circuit optimizer~\cite{VOQC}.
In this paper, we presented \sqir as a source language for quantum programming and discussed how our design choices (e.g. concrete indices, unitary core, vector state abstractions) ease proofs about \sqir programs.
But there is still work to be done.

So far, work on formally verified quantum computation has been limited to textbook quantum algorithms like QPE and Grover's.
Although these algorithms are a useful stress-test for tools, they do not accurately reflect the types of quantum programs that are expected to run on near-term machines.
Near-term algorithms are usually \emph{approximate}.
They do not implement the desired operation exactly, but rather perform an operation ``close'' to what was intended.
Our \coqe{probability_of_outcome} and \coqe{prob_partial_meas} predicates can be used to express distance between vector states, but we currently do not have support for reasoning about distance between general quantum operations.

Another issue is that near-term algorithms often need to account for hardware errors.
Thus, verifying these algorithms may require considering their behavior in the presence of errors.
So far, most of our work in \sqir has revolved around the unitary semantics and vector-based state abstractions because we find these simpler to work with.
However, it is more natural to describe states subject to error using density matrices, since noisy states are mixtures of pure states \cite[Chapter 8]{NCbook}.

On another front, there is important work to be done on describing quantum algorithms and correctness properties at a higher level of abstraction. 
The proofs and definitions in this paper follow the standard textbook presentation, but are still lower-level than similar proofs about classical programs.
Rather than working from the circuit model, used in 
\qwire, \sqir, \qbricks, and (to some extent) QWhile, it would be interesting to verify programs written in higher-level languages like Silq~\cite{Bischel2020} or Q\#~\cite{Svore2018}.

We hope that \sqir's extensible design and flexible semantics, developed while verifying circuit optimizations and textbook quantum programs, will serve as a solid foundation for the proposed verification efforts above and those to come.

\bibliography{references}

\clearpage
\appendix


\section{SQIR with Measurement}
\label{app:measurement}

Full \sqir extends the unitary fragment with support for measurement. The syntax for full \sqir and its density matrix-based semantics were discussed in \Cref{sec:measurement}. Here we present an alternative, \emph{nondeterministic} semantics and an example \sqir program with measurement. 

\subsection{Nondeterministic Semantics}

In addition to the density matrix-based semantics described in \Cref{sec:measurement}, \sqir also supports a \emph{nondeterministic semantics} in which evaluation is expressed as a relation.
Given a state $\ket{\psi}$, a unitary program \coqe{u} will (deterministically) evaluate to \coqe{[[u]]${}_d$ $\times \ket{\psi}$}.
However, \coqe{meas q p1 p2} may evaluate to either \coqe{p1} applied to $\op{1}{1} \times \ket{\psi}$ or \coqe{p2} applied to $\op{0}{0} \times \ket{\psi}$.
We use notation \coqe{p / $\psi$ ⇩ $\psi'$} to say that on input $\psi$ program \coqe{p} nondeterministically evaluates to $\psi'$.

The advantage of the nondeterministic semantics is that state is represented using a vector $\ket{\psi}$ rather than a density matrix $\rho$, which makes proofs easier (\Cref{sec:unitarycore}).
However, because the nondeterministic semantics only describes one possible measurement outcome, it is only useful for proving certain types of properties.
For example, it can be used to prove the existence of a possible output state or to show that all execution paths result in the same outcome. The following examples share the latter property, allowing us to compare the density matrix-based and nondeterministic semantics.

\paragraph*{Example: Resetting a Qubit}

Consider the following \sqir program, which resets qubit \coqe{q} to the $\ket{0}$ state.
\begin{coq}
Definition reset q = meas q (X q) skip.
\end{coq}
Using our density matrix-based semantics, we can prove the following, which says that for any valid density matrix $\rho$, applying \coqe{reset} to $\rho$ will produce the density matrix corresponding to the $\ket{0}$ state.
\begin{coq}
Lemma reset_to_zero: forall (\rho : Density 2), Mixed_State \rho -> [[[reset]]]$_1$ \rho = \k0\b0.
\end{coq}
The proof is straightforward:
\begin{align*}
\denote{\mathtt{reset}}~\rho 
&= X(\ketbra{1}{1}\rho\ketbra{1}{1})X + I_2(\ketbra{0}{0}\rho\ketbra{0}{0})I_2 \\
&= \ketbra{0}{1}\rho\ketbra{1}{0} + \ketbra{0}{0}\rho\ketbra{0}{0} \\
&= \ket{0}(\bra{1}\rho\ket{1} + \bra{0}\rho\ket{0})\bra{0} = \ket{0}(I_1)\bra{0} = \ketbra{0}{0}
\end{align*}
The last line uses the fact that $\rho$ is a valid density matrix (\coqe{Mixed_State}), which implies that the entries along its diagonal sum to $1$. 

Although the proof above is straightforward, it does not give a clear intuition for \emph{why} the program is correct.
The simple explanation for why this program is correct is as follows: There are two cases, depending on the result of \coqe{meas}. In the case where measurement outputs 0, the remainder of the program is the no-op \coqe{skip}, so the output state is $\ket{0}$. In the case where measurement outputs 1, the program applies an $X$ gate, which flips the qubit's value, leaving it in final state $\ket{0}$.

The proof using the nondeterministic semantics closely follows this argument: It considers both possible measurement transitions and inspects the output state. The correctness property for the nondeterministic semantics is stated as follows.
\begin{coq}
Lemma reset_to_zero: forall ($\psi$ $\psi'$ :Vector 2), WF_Matrix $\psi$ -> reset / $\psi$ ⇩ $\psi'$ -> $\psi'$ ∝ ∣0⟩
\end{coq}
This says that \emph{any} output state $\psi'$ is proportional ($\propto$) to $\ket{0}$. 


\subsection{Quantum Teleportation}
\label{sec:teleport}

The goal of quantum teleportation is to transmit a state $\ket{\psi}$ from one party (Alice) to another (Bob) using a shared entangled state. The circuit for quantum teleportation is shown in \Cref{fig:teleport-circ} and the corresponding \sqir program is given below.
\begin{coq}
Definition bell : ucom base 3 := H 1; CNOT 1 2.
Definition alice : com base 3 := CNOT 0 1 ; H 0; measure 0; measure 1.
Definition bob : com base 3 := CNOT 1 2; CZ 0 2; reset 0; reset 1. 
Definition teleport : com base 3 := bell; alice; bob.
\end{coq}
The \coqe{bell} circuit prepares a Bell pair on qubits 1 and 2, which are respectively sent to Alice and Bob. Alice applies \coqe{CNOT} from qubit 0 to qubit 1 and then measures both qubits and (implicitly) sends them to Bob. Finally, Bob performs operations controlled by the (now classical) values on qubits 0 and 1 and then resets them to the zero state.

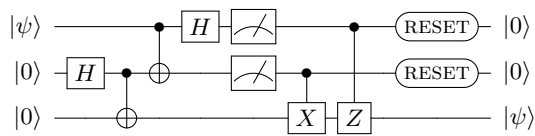
\begin{figure}[t]
\centering
\centerline{
  \small
  \Qcircuit @C=0.5em @R=0.5em {
    \lstick{\ket{\psi}} & \qw & \qw & \ctrl{1} & \gate{H} & \meter & \qw & \ctrl{2} & \qw & \measure{\mbox{\textsc{reset}}} & \rstick{\ket{0}} \qw \\
    \lstick{\ket{0}} & \gate{H} & \ctrl{1} & \targ & \qw  & \meter & \ctrl{1} & \qw & \qw & \measure{\mbox{\textsc{reset}}} & \rstick{\ket{0}} \qw\\
    \lstick{\ket{0}} & \qw & \targ & \qw & \qw & \qw & \gate{X} & \gate{Z} & \qw & \qw & \rstick{\ket{\psi}} \qw
    }
}
  \caption{Circuit for quantum teleportation following our presentation in prior work \cite[Sec. 5]{VOQC}.}
  \label{fig:teleport-circ}
\end{figure}

\paragraph*{Density Matrix-Based Semantics}
The correctness property for this program says that for any (well-formed)
density matrix $\rho$, \coqe{teleport} takes the state $\rho \otimes \op{0}{0} \otimes \op{0}{0}$ to the state $\op{0}{0} \otimes \op{0}{0} \otimes \rho$.
\begin{coq}
Lemma teleport_correct : forall (ρ : Density 2),
 WF_Matrix ρ -> 
 $\pdenote{\text{teleport}}_3$ (ρ ⊗ ∣0⟩⟨0∣ ⊗ ∣0⟩⟨0∣) = ∣0⟩⟨0∣ ⊗ ∣0⟩⟨0∣ ⊗ ρ
\end{coq}
The proof is simple: We perform (automated) arithmetic to show that the output matrix has the desired form.

\paragraph*{Nondeterministic Semantics}
Under the nondeterministic semantics, we aim to prove the following, which says that on input $\ket{\psi} \otimes \ket{0,0}$, \coqe{teleport} will produce a state that is proportional to $\ket{0,0} \otimes \ket{\psi}$.
\begin{coq}
Lemma teleport_correct : forall (ψ : Vector (2^1)) (ψ' : Vector (2^3)),
  WF_Matrix ψ -> teleport / (ψ  ⊗ ∣0,0⟩) ⇩ ψ' -> 
  ψ' ∝ ∣0,0⟩ ⊗ ψ.   
\end{coq}
The first half of the circuit is unitary, so the proof simply computes the effect of applying a $H$ gate, two $\mathit{CNOT}$ gates and another $H$ gate to the input vector state. The two measurement steps then leave four different cases to consider. In each of the four cases, we can use the outcomes of measurement to correct the final qubit, putting it into the state $\ket{\psi}$. Finally, resetting the already-measured qubits is deterministic and leaves us with the desired state.

\section{Additional Examples}
\label{app:general-verification}

Here we present three additional quantum algorithms we have verified in \sqir. We begin with \emph{superdense coding}, a simple algorithm whose proof is almost entirely automated. We then comment on the \emph{Deutsch-Jozsa algorithm}, the first algorithm we verified that is parameterized by an input oracle (Grover's algorithm and QPE both share this property). We conclude with \emph{Simon's algorithm}, another parameterized program similar in structure to Deutsch-Jozsa.

\subsection{Superdense Coding}
\label{sec:superdense}

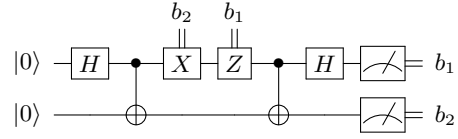
\begin{figure}[t]
\begin{subfigure}{0.5\textwidth}
\begin{coq}
Definition bell00 := H 0; CNOT 0 1.
Definition encode (b1 b2 : bool) := 
  (if b2 then X 0 else I 0);
  (if b1 then Z 0 else I 0).
Definition decode := CNOT 0 1; H 0.
Definition superdense (b1 b2 : bool) := 
  bell00 ; encode b1 b2 ; decode.
\end{coq}
\end{subfigure}
\begin{subfigure}{0.5\textwidth}
\centerline{
  \small
  \Qcircuit @C=0.7em @R=0.7em {
     &  &  & b_2 & b_1 & & & & \\
     &  &  & \cwx[1] & \cwx[1] & & & & \\
    \lstick{\ket{0}} & \gate{H} & \ctrl{1} & \gate{X} & \gate{Z} & \ctrl{1} & \gate{H} & \meter & \rstick{b_1} \cw \\
    \lstick{\ket{0}} & \qw & \targ & \qw & \qw & \targ & \qw & \meter & \rstick{b_2} \cw
    }
}  
\end{subfigure}
\vspace{-1em}
\caption{Superdense coding in \sqir and as a circuit. Each definition in the \sqir program has type \coqe{ucom base 2}.}
  \label{fig:superdense}
\end{figure}

Superdense coding is a protocol that allows a sender to transmit two classical bits, $b_1$ and $b_2$, to a receiver using a single quantum bit. Initially, the sender and receiver share an entangled pair of qubits called a \emph{Bell pair}. To start, the sender conditionally applies $X$ and $Z$ to their qubit, contingent on the values of $b_1$ and $b_2$, and then transmits their qubit to the receiver. The receiver then applies a \emph{Bell measurement} (reversed entangling operation followed by measure) to recover the bits.
The circuit for superdense coding and the \sqir program corresponding to the unitary part of this circuit are shown in \Cref{fig:superdense}.
   
We can prove that the result of evaluating the program \coqe{superdense b1 b2} on an input state consisting of two qubits initialized to zero is the state $\vert b_1, b_2 \rangle$.
\begin{coq}
Lemma superdense_correct : forall b1 b2, [[superdense b1 b2]]${}_2$ $\times$ $\vert$ 0,0 $\rangle$ = $\vert$ b1,b2 $\rangle$.
\end{coq}
The proof simply destructs \coqe{b1} and \coqe{b2} and applies matrix simplification tactics.

\subsection{Deutsch-Jozsa Algorithm}

In the Deutsch-Jozsa problem~\cite{deutsch1992rapid}, the goal is to determine whether a Boolean function $f:\{0,1\}^n\to\{0,1\}$ is \emph{constant} (always returns the same value) or \emph{balanced} (returns 0 and 1 equally often), given that one is the case. 
The function $f$ is encoded in an ``oracle'' $U_f:\ket{x,y}\mapsto\ket{x,y\oplus f(x)}$, which is a linear operator over a $2^{n+1}$ dimensional Hilbert space.
In \sqir, we express the requirement that program \coqe{U_f} encodes the function $f$ as follows.
\begin{coq}
Definition boolean_oracle {n} (U_f : ucom (n + 1)) f :=
  forall x y, [[u]]$_{n+1}$ × ∣x⟩ ⊗ ∣y⟩ = ∣x⟩ ⊗ ∣y ⊕ (f x)⟩.
\end{coq}
To express that a function is constant or balanced, we can define a function \coqe{count f n} that counts all inputs on which function \coqe{f} (with domain size $2^n$) evaluates to true. Then we have:
\begin{coq}
Definition balanced f n := n > 0 /\ count f n = $2 ^ {n - 1}$.
Definition constant f n := count f n = 0 \/ count f n = $2 ^ n$.
\end{coq}

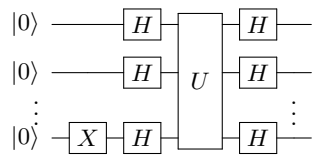
\begin{figure}[t]
\begin{subfigure}{0.6\textwidth}
\begin{coq}
Definition deutsch_jozsa n (U_f : ucom base n) :=
  X (n-1) ; npar n H ; u ; npar n H.
\end{coq}
\end{subfigure}
\begin{subfigure}{0.05\textwidth}
~
\end{subfigure}
\begin{subfigure}{0.3\textwidth}
\vspace{1em}
\centerline{
  \small
  \Qcircuit @C=0.7em @R=0.7em {
     \lstick{\ket{0}} & \qw & \gate{H} & \multigate{3}{U} & \gate{H} & \qw & \qw \\
     \lstick{\ket{0}} & \qw & \gate{H} & \ghost{U} & \gate{H} & \qw & \qw \\
     \lstick{\vdots} &  &  &  &  & \vdots &  \\
     \lstick{\ket{0}} & \gate{X} & \gate{H} & \ghost{U} & \gate{H} & \qw & \qw \\
    }
}  
\end{subfigure}
  \caption{The Deutsch-Jozsa algorithm in \sqir and as a circuit. The Coq function \coqe{npar} constructs a \sqir program that applies the same operation to every qubit. }
  \label{fig:deutsch}
\end{figure}

As shown in \Cref{fig:deutsch}, the Deutsch-Jozsa algorithm begins with an all $\ket{0}$ state and prepares the input state \coqe{∣+)$^{\otimes n}$ ⊗ ∣-⟩} by applying an $X$ gate on the last qubit followed by a $H$ gate on every qubit.
Next the oracle \coqe{U_f} is evaluated, and a $H$ gate is again applied to every qubit in the program.
Finally, all qubits are measured in the standard basis.
If measuring all qubits but the last yields an all-zero string (the last qubit is guaranteed to be in the $\ket{1}$ state) then the algorithm outputs ``accept,'' indicating that the function is constant.
Otherwise the algorithm outputs ``reject.''

The probability of measuring $\ket{0}$ in the first $n$ qubits is given by 

\coqe{        prob_partial_meas $\ket{0}^{n}$ ([[deutsch_jozsa n U_f]]$_{n+1}$ × ∣0⟩$^{n+1}$)}.

We define an \coqe{accept} predicate that states that this expression is 1 and a \coqe{reject} predicate that states that it is 0.
The correctness property is then stated as follows.
\begin{coq}
Lemma deutsch_jozsa_correct : forall (n : nat) (f : nat -> bool) (U_f : ucom base (n + 1)), 
  n > 0 -> boolean_oracle u f -> 
  (constant f n -> accept U_f) /\ (balanced f n -> reject U_f).
\end{coq}

The key lemma in our proof states that the probability of accepting depends on the number of inputs on which \coqe{f} evaluates to 1, i.e., \coqe{count f n}. In particular, the probability is $Pr_{\mathit{accept}} = |1 - \frac{2 * (\text{\coqe{count f n}})}{2 ^ n}|^2$.
We prove this using matrix simplification and induction on $n$.
Correctness of the Deutsch-Jozsa algorithm follows directly from this lemma:
For a constant function, $\text{\coqe{count f n}} = 0$ or $\text{\coqe{count f n}} = 2^n$ so $Pr_{\mathit{accept}} = 1$. For a balanced function, $\text{\coqe{count f n}} = 2 ^ {n - 1}$ so $Pr_{\mathit{accept}} = 0$.

\subsection{Simon's Algorithm}

Given a function $f: \{0,1\}^n\to\{0,1\}^n$ such that for all $x, y \in \{0,1\}^n$, $f(x)=f(y)\Leftrightarrow x\oplus y \in \{0,s\}$ for unknown $s \in \{0,1\}^n$, the goal of Simon's algorithm is to find $s$. 
The inputs to the algorithm are the input size $n$ and a program (oracle) $U_f$ with the property that $U_f\ket{x}\ket{y}=\ket{x}\ket{f(x)\oplus y}$.
If $s=0$, then the output of Simon's algorithm is a uniform superposition over all $n$-bit strings (meaning that any string is measured with equal probability). If $s\neq 0$, then the output is a uniform distribution over strings $y$ such that $s \cdot y = 0$, where $x \cdot y$ is the bitwise dot product of $x$ and $y$ modulo 2. The value of $s$ can be determined by $O(n)$ iterations of the algorithm.

The \coqe{simon} function, shown below, produces the \sqir circuit for the algorithm, which has a structure similar to Deutsch-Jozsa. First, a layer of Hadamard gates prepares a uniform superposition on the first $n$ inputs. Next, $U_f$ encodes information about $f$ in the phase, in essence evaluating the oracle on all possible inputs at once. Finally, another layer of Hadamard gates brings information in the phase back to the state where it can be measured. The circuit is run on input $\ket{0}^{2 * n}$ ($= \ket{00...0}$ with $2 * n$ entries).
\begin{coq}
Definition simon {n} (Uf : ucom base (2 * n)) := npar n H ; Uf ; npar n H.
\end{coq}

Our statements of correctness for Simon's algorithm say that (1) if $s$ is zero then the probability of measuring any particular output is $1/2^n$, (2) if $s$ is nonzero then the probability of measuring $y$ such that $s \cdot y = 0$ is $1/2^{n-1}$, and (3) if $s$ is nonzero then the probability of measuring $y$ such that $s \cdot y \neq 0$ is 0. We show the full statement of correctness for property (2) below.

\begin{coq}
Lemma simon_nonzero_A : forall {n : nat} (Uf : ucom base (2 * n)) f y s,
  n > 0 -> y < $2^n$ -> s < $2^n$ ->
  integer_oracle Uf f ->
  (forall x, x < $2^n$ -> f x < $2^n$) ->
  (forall x y, x < $2^n$ -> y < $2^n$ -> f x = f y <-> $x \oplus y$ = s \/ x = y)) ->
  s <> 0 ->
  $s \cdot y$ = 0 ->
  prob_partial_meas $\ket{y}$ ([[simon Uf]]$_{2 * n}$ × $\ket{0}^{2 * n}$) = $\frac{1}{2^{n-1}}$.
\end{coq}
The first three conditions ensure well-formedness of the inputs; the next three describe constraints on $f$ and state that \coqe{Uf} implements $f$. 
We call \coqe{Uf} an \emph{integer oracle} because it maps an $n$-bit number to another $n$-bit number.
The conclusion states that after running the program \coqe{simon Uf} on $\ket{0}^{2*n}$, the probability of measuring $y$ such that $s \cdot y = 0$ is $\frac{1}{2^{n-1}}$.

We begin by showing that for any (well-formed) $s$ and $y$, 
\begin{center}
    \coqe{prob_partial_meas $\ket{y}$ ([[simon Uf]]$_{2 * n}$ × $\ket{0}^{2 * n}$)} $= \lVert \frac{1}{2^n} \sum_{x=0}^{2^n - 1}(-1)^{x \cdot y} \ket{f(x)} \rVert^2.$
\end{center}
The proofs of the three properties listed above then amount to showing properties about this norm-sum term.

In the case where $s \neq 0$, $f$ is a two-to-one function, which means that the expression above can be rewritten as a sum over elements in the range of $f$. In the standard presentation, this expression is simplified as follows.
\begin{align*}
\small
&\big\lVert \frac{1}{2^n} \sum_{z \in \text{range}(f)}((-1)^{x_1 \cdot y} + (-1)^{x_2 \cdot y}) \ket{z} \big\rVert^2, \qquad f(x_1) = f(x_2) = z \\
&= \big\lVert \frac{1}{2^n} \sum_{z \in \text{range}(f)}((-1)^{x_1 \cdot y} + (-1)^{(x_1 \oplus s) \cdot y}) \ket{z} \big\rVert^2 \\
&= \big\lVert \frac{1}{2^n} \sum_{z \in \text{range}(f)}((-1)^{x_1 \cdot y}(1 + (-1)^{s \cdot y}) \ket{z} \big\rVert^2
\end{align*}
From this rewritten form, it is clear that the probability of measuring $y$ such that $s \cdot y = 0$ is $2 * 1/2^n = 1/2^{n-1}$ and the probability of measuring $y$ such that $s \cdot y \neq 0$ is $0$.

Our Coq proof essentially follows this structure, although we found it easier to define a function \coqe{to_injective} that takes the two-to-one function $f$ and makes it one-to-one.
\begin{coq}
Definition to_injective n s f x :=
  let y := $x \oplus s$ in 
  if (x <? y) then f x else ($2 ^ n$ + f x).
\end{coq}
Using this function, we can rewrite the norm-sum term as a sum over vectors of size $2^{n+1}$.
\[
    \big\lVert \sum_{x=0}^{2^n - 1} (-1)^{x \cdot y}\ket{f(x)} \big\rVert = \frac{1}{\sqrt{2}} \big\lVert \sum_{x=0}^{2^n - 1} ((-1)^{x \cdot y} + (-1)^{(x \oplus s) \cdot y}) \ket{(\small\texttt{to\_injective~n~s~f})(x)} \big\rVert
\]


\end{document}